\documentclass[12pt]{article}
\usepackage{amsmath,amsthm,amssymb,euscript,epsf,epsfig,color}
\usepackage{array}
\usepackage{graphicx}
\usepackage{fancybox}
\usepackage[font=small,labelfont=bf]{caption}
\usepackage{subcaption}
\usepackage{flafter}
\usepackage{comment}
\graphicspath{{/Users/Lewissword/}}

\newcommand{\be}{\begin{equation}}
\newcommand{\ee}{\end{equation}}
\newcommand{\bea}{\begin{eqnarray}\displaystyle}
\newcommand{\eea}{\end{eqnarray}}

\makeatletter
\@addtoreset{equation}{section}
\makeatother
\renewcommand{\theequation}{\thesection.\arabic{equation}}
\def\one{{\hbox{ 1\kern-.8mm l}}}
\def\zero{{\hbox{ 0\kern-1.5mm 0}}}

\def\cG{{\cal G}}  
  \def\cL{{\cal L}}
  \def\cO{{\cal O}}

 \def\cZ{{\cal Z}}

\begin{document}

\makeatletter
\@addtoreset{equation}{section}
\makeatother
\renewcommand{\theequation}{\thesection.\arabic{equation}}

\rightline{QMUL-PH-19-29}
\vspace{1.8truecm}

\vspace{10pt}


{\LARGE{ 
\centerline{\bf  Gaussianity and typicality in  } 
\centerline{\bf   matrix  distributional semantics } 
}}  

\vskip.5cm 

\thispagestyle{empty} \centerline{
   {\large \bf  Sanjaye Ramgoolam${}^{a,b,}$\footnote{ {\tt s.ramgoolam@qmul.ac.uk}},    }
               {\large \bf  Mehrnoosh Sadrzadeh${}^{c,}$\footnote{ {\tt m.sadrzadeh@ucl.ac.uk }}   
               and {\large \bf Lewis Sword ${}^{a,}$\footnote{ {\tt l.sword@se15.qmul.ac.uk }}. } }
                                                       }

\vspace{.4cm}
\centerline{{\it ${}^a$ Centre for Research in String Theory, School of Physics and Astronomy},}
\centerline{{ \it Queen Mary University of London, Mile End Road, London E1 4NS, UK. }}
    
    \vspace{.2cm}
\centerline{{\it ${}^b$ School of Physics and Centre for Theoretical Physics, }}
\centerline{{\it University of the Witwatersrand, Wits, 2050, South Africa } }

    \vspace{.2cm}
\centerline{{\it ${}^c$ University College London, Department of Computer Science,}}
\centerline{{\it  Gower Street, London WC1E 6BT, United Kingdom  }}

\vspace{.5truecm}

\thispagestyle{empty}

\centerline{\bf ABSTRACT}

\vskip.2cm 

Constructions in type-driven compositional distributional semantics associate 
large collections of  matrices of size $D$  to linguistic corpora. We develop the proposal of analysing the statistical characteristics of this data in the framework of permutation invariant matrix models. The observables in this framework are permutation invariant polynomial functions of the matrix entries, which correspond to directed graphs. Using the general 13-parameter permutation invariant 
Gaussian matrix models recently solved, we find, using a dataset of matrices constructed via 
standard techniques in distributional semantics, 
 that the expectation values of  a large class of cubic and quartic observables show high gaussianity at levels between 90 to 99 percent. Beyond expectation values, which are averages over words, the dataset allows the computation of standard deviations for each observable, which can be viewed as a measure of typicality for each observable.  There is a wide range of magnitudes in the measures of typicality.   The permutation invariant matrix models, considered as functions of random couplings, give a very good prediction of the magnitude of the typicality for different observables. We find evidence that observables with similar matrix model characteristics of Gaussianity and typicality also have  high degrees of correlation between the ranked lists of words  associated to these observables.

\setcounter{page}{0}
\setcounter{tocdepth}{2}

\newpage

\tableofcontents

\section{ Introduction}

A  research programme ``Linguistic Matrix Theory'' of understanding the characteristics of randomness in natural language, specifically in matrix/tensor datasets arising  from type-driven compositional distributional semantics, using  the framework of random matrix/tensor  theories was initiated in \cite{LMT}. 

Distributional or vector space models of meaning in natural language semantics argue that meanings of words are representable by the contexts in which they often occur. The ideas that have inspired this way of reasoning about meaning go back to the works of Firth \cite{JRF57} and of Harris \cite{ZH68}, the former of which famously said: ``You shall  know the meaning of a word by the company  it keeps."   These ideas were  implemented via  vectors  of co-occurrence contexts \cite{GS75,RG65}. Contexts, e.g. words in a fixed neighbourhood window of size $k$, are taken to be the basis of a vector space whose elements represent meanings of other words. The coefficients of a word vector  $w$ over a basis vector  $b_i$  is a function of the number of times $w$ occurred in the context of $b_i$. These co-occurrence frequencies are collected from  large corpora of data, such as crawls of web domains, Google's news and book archives, and the  Wikipedia. The distances between the word vectors represent their semantic similarity/relatedness, e.g. see \cite{HS98}. 

 In order to extend the distributional model from words to phrases and sentences, one has to take into account  grammatical structure. Type-logical approaches to grammar, e.g. Combinatory Categorial Grammar \cite{MS2001} and the Lambek Calculus \cite{JL58},  have been shown to have a straightforward interface to the vector space models of meaning.  The ideas behind these grammatical formalisms are the same, although they follow different notational conventions and syntactic  rules and in this paper we adopt the terminology of Lambek Calculus.  In a type-logical grammar,    some words, such as nouns, have  atomic types and others, such as adjectives and verbs, have functional (or functor) types. If we start with the set $\{n,s\}$ of atomic types, $n$ for the type of a noun and $s$ for the type of a sentence, then an adjective will have type $n \to n$: this type says that an adjective is a function that takes an argument of type noun, modifies it and returns an adjective  noun phrase of type $n$. For instance, the adjective ``red" takes  the noun ``cat" as an argument and after modifying it,  returns the phrase ``red cat" as an adjective noun phrase. An intransitive verb has type $n \to s$, i.e. a function that takes an argument of type noun and returns a sentence. An example here is the  verb ``snore", which takes the noun ``cats" as argument and returns the sentence ``cats snore". A transitive verb has type $n \to n \to s$; this is a function that takes an arguments of type noun, returns a verb phrase of type $n \to s$, which in turn takes an argument of type noun and returns a sentences. An example is the verb ``like", it takes the noun ``fish" and returns the verb phrase `like fish", which in turn takes the noun ``cats" and returns the sentence ``cats like fish"\footnote{One can think of a transitive verb as a function of two nouns and thus assign the type $n \times n \to s$ to it. Following Chomsky, however, a sentence must be generated by a noun phrase followed by a verb phrase and despite the equivalence $n \to n \to s \equiv n \times n \to s$, syntactically these two types should be distinguished. }.

Type-driven  distributional models of meaning start from a type-driven analysis of grammar and assign a compositional vector semantics to  natural language constructions  \cite{CCS2010,SC2013,CGS2013,MCG2014}. These models  are based on the  centreline argument that words with atomic types should be  represented as vectors, but words with  functional types  as linear or multilinear maps, equivalently matrices, cubes, or higher order tensors, depending on the number of  arguments they take. If we assign the vector space $N$ to the type $n$ and the vector space $S$ to the type $s$,  then adjectives,  intransitive verbs, and verb phrases that only have one argument become elements of the tensor spaces $N \otimes N$ and $N \otimes S$, respectively. These  are represented as matrices. Transitive and ditransitive verbs have  two and three arguments each: they become elements of the tensor spaces $N \otimes N \otimes S$ and $N \otimes N \otimes N \otimes S$ and  are represented as cubes and hypercubes, respectively. When a word with a functional type composes with a word with an atomic type, the composition is represented by the application of the corresponding linear/multilinear map with the vector of the atomic word. As an example, consider  the distributional  meaning of  ``red cat", which becomes the result of the matrix multiplication of the  matrix of ``red" and  vector of ``cat".  Denoting the former with $M^{red}_{ij} \in N \otimes N$ and the latter with $V^{cat}_j \in N$,  we obtain $M^{red}_{ij} \times V^{cat}_j$ as the meaning of ``red cat". Similarly, denoting the meaning of ``snore" by $M^{snore}_{ij} \in N \otimes S$, we obtain $M^{snore}_{ij} \times V^{cats}_j$ as the meaning of ``cats snore". Similarly, the meaning of ``cats like fish"  becomes  $M^{like}_{ijk} \times V^{fish}_k \times V^{cats}_j$ and so on.

The  collection of matrices associated to  adjectives and intransitive verbs of a corpus 
have large  matrix sizes,  ranging from $100$ up to  $10$K, and   to $40$K, e.g. see the original work of \cite{BZ2010} for the higher dimensions  and the sequel  work of \cite{MC2015} for the lower ones. While the AI inspired tasks are focused on 
extracting linguistic structure, e.g. word similarity, from these matrices,  such a large collection inevitably has elements of randomness. Any corpus is a finite, even if large,  sample selected from everything written in a language. Even if it is a good approximation to everything written, the written words in a corpus are influenced by the experience of the authors, subject for example to a wide range of interactions with the environment and other humans. We may ask if there are universal patterns in the randomness existing in the large datasets of matrices encoding the complex natural system that is human language. The experience of random matrix theory has indeed shown that the patterns in the distribution of energy eigenvalues of complex nuclei \cite{Wigner,Dyson} also occur in a wide variety of complex systems (see for example \cite{GMW98,Been1997,EY2013,RMTMN0503}).

In  the Linguistic Matrix Theory (LMT)  programme of \cite{LMT}, one of the first  
steps was to identify the appropriate type of symmetry. Here it was useful to consider the kinds of mathematical expressions which are used in distributional semantics to extract the meaning encoded in words. For vector, matrix and tensor data in $D$ dimensions, some of these expressions are invariant under the orthogonal group of all rotations in $D$ dimensions, but the generic expressions are only invariant under the smaller symmetry of all permutations of $D$ objects, the symmetric group $S_D$. This motivated us to consider matrix models with $S_D$ symmetry. The polynomial functions of matrix variables $M_{ ij}$ which are $S_D$ invariant have an elegant classification in terms of polynomials labelled by directed graphs. The degree of the polynomial is the number of edges in the graph : the number of nodes is unconstrained. There are two graphs at linear order, each associated with a permutation invariant polynomial. A general permutation invariant linear function is a sum of these two polynomials with arbitrary coefficients. We restrict these linear coefficients to be real numbers numbers $ \mu_1 , \mu_2 $. There are eleven independent quadratic functions. As a simple toy model we considered 
three quadratic polynomials with three associated coefficients $ \Lambda_1 , \Lambda_2 , \Lambda_3 $. 
We defined a function $ S ( \mu_1 , \mu_2 , \Lambda_1 , \Lambda_2 , \Lambda_3 )$ and considered a probability distribution defined by the partition function  
\bea 
\cZ = \int dM  e^{ - S ( \mu_1 , \mu_2 , \Lambda_1 , \Lambda_2 , \Lambda_3 ) } 
\eea
Given any permutation invariant polynomial, which we will henceforth refer to as observables and denote $ \cO ( M ) $, we 
can calculate a theoretical expectation value 
\bea
\langle \cO \rangle_{ THEO } = {1 \over \cZ } 
\int dM  e^{ - S ( \mu_1 , \mu_2 , \Lambda_1 , \Lambda_2 , \Lambda_3 ) } \cO ( M )  
\eea
The  expectation values of the linear and quadratic observables  $ \langle \cO  \rangle $ are expressible as simple functions of the $ \mu_a , \Lambda_i $. In order to match these probability distributions 
with experimental data, the experimental expectation values for these five observables were computed 
as averages over the words in the dataset 
\bea 
{ 1 \over N_{words}   } \sum_{ A } \cO ( M^A  ) 
\eea
$A$ is a label for the words in the dataset and $N_{words} $ is the number of words in the dataset. 
Equating these to the theoretical expectation values, we determined the $ \mu_a , \Lambda_i $ parameters of the model, for a given dataset. 

The theoretical model was also used to calculate the expectation values of a number of cubic and quadratic observables. These theoretical values, using the input of $ \mu_a, \Lambda_i $ determined as above, 
give the predictions of the 5-parameter Gaussian model for these observables. We calculated the ratios of 
the theoretical to experimental values, with a ratio close to $1$ being good agreement between theory and experiment. The best ratios were approximately 60 \%, but for a number of observables the ratios were very low, the lowest being around 0.6 \% . We argued that a more complete treatment with a general Gaussian model that includes all the eleven parameters would likely give better ratios. 

The theoretical model with all eleven quadratic parameters was solved in \cite{PIGMM}. 
It was useful to employ a representation theoretic approach to the space of quadratic permutation invariant functions. The eleven parameters were organised according to four irreducible representations 
$ V_0 , V_H , V_2 , V_3 $ of $S_D$. $ \Lambda^{ V_0 } $ is a symmetric $ 2 \times 2 $ matrix with three real parameters, $ \Lambda^{ V_H } $ is a symmetric $ 3 \times 3 $ real matrix with $6$ parameters and $ \Lambda^{ V_2 } , \Lambda^{ V_3} $ are each real numbers. We have an action 
\bea 
 S ( \mu_a , \Lambda^{ V_0 } , \Lambda^{ V_H} , \Lambda^{ V_2 } , \Lambda^{ V_3 } ) \equiv 
S ( \mu_a , \Lambda^V )  
\eea
 which defines a probability distribution and associated partition function 
\bea 
\cZ = \int dM   e^{ - S ( \mu_a , \Lambda^V ) } 
\eea 
Convergence of the measure requires that $ \Lambda^{ V_0} , \Lambda^{ V_H}  $ are
 positive semi-definite matrices, and $ \Lambda^{ V_2} \ge 0 , \Lambda^{ V_3} \ge 0$.   
The first main  goal of this paper is to report on the application of
 this 13-parameter Gaussian model from \cite{PIGMM} to the same dataset constructed in \cite{LMT}, to test its effectiveness at predicting cubic and  quartic expectation values along the lines of the approach in \cite{LMT}. 
 
It is useful at this point  to discuss our investigations of Gaussianity in language in the broader context of studying statistical aspects of language and of applications of matrix models in physics. 
The two central elements of the programme initiated in \cite{LMT} are Gaussianity and permutation symmetry. It is worthwhile discussing  a potential objection to Gaussianity in linguistics. 
Zipf's Law \cite{Zipf,WLi}  is the observation that in corpora of natural language, e.g. collections of written text in
a language such as English, the frequency of a word is inversely proportional to its rank. A power law is of course nothing like a Gaussian, so a quick argument is that Gaussians are not what one typically gets in the statistics of  linguistic corpora.

Since the 1990's, starting from the works \cite{BK,GM,DS},  matrix theories have seen an explosion of applications in theoretical physics, to match   their widespread use in applied physics along the lines of the original work of Wigner and Dyson. A prominent area of application is gauge-string duality, where quantum field theories with matrix degrees of freedom in diverse dimensions, have an emergent dual description in terms of a string theory. The present application of matrix models to language, can be viewed as the exploration of another instance of emergence from matrix theory. In the present case, the emergence is that of universal aspects of randomness in  natural language from the mathematics of matrix theory. 

One of the lessons from the applications of matrix models in gauge-string duality and indeed more broadly from  applications of quantum field theory,  is that the validity of Gaussianity or approximate Gaussianity often needs to be carefully delineated in a complex system.  Following the analogy between integrals and path integrals describing quantum field theories, field theoretic realizations of Gaussianity or near-Gaussianity are phenomena described by free quantum field theories where the actions are quadratic in the field variables, and  perturbations of these free theories by small higher order corrections.  Taking the field theory to be the gravitational space-time field theory in the AdS string background in the context of the AdS/CFT correspondence \cite{malda}, an important instance of this is the failure of perturbative gravity in accounting for high energy graviton interactions \cite{BBNS} even at a qualitative level, which is related to the phenomenon of giant gravitons \cite{MST}. In the dual matrix CFT description of this physics, this giant graviton physics appears  from the combinatorics of large composite operators associated with particular shapes of Young diagrams \cite{CJR}. The Gaussian regime of perturbative gravitons arises from correlators of low order gauge invariant polynomials. The lesson is that the phenomena described  by matrix theories are rich and diverse, with Gaussianities emerging from the identification of the appropriate observables. 

The same theme is evident in the applications of quantum field theory to the real world. In the context of quantum  field theories applied to particle physics phenomena, free-field behaviour (approximate Gaussianity)  arises in the high energy (ultra-violet) regime for theories such as quantum chromodynamics (QCD) while it arises  in the low-energy (infra-red) regime for quantum electrodynamics
(see for example textbooks in quantum field theory such as \cite{Schwarz}).  In  cosmology, the detailed study of the approximate Gaussianity of  fluctuations in the cosmic microwave background, which originate from a very early period in the history of the universe,  and bounds on the non-Gaussianities are used to constrain the space of theoretical models of inflation \cite{Planck2015}. In this setting, the fluctuations of the temperature of the CMB in different directions in the sky are experimentally measurable observables which can be related to theoretical models of inflation described by a path integral for a scalar field. In the present context of type-driven compositional distributional semantics, the experimental data consists of matrices constructed from linguistic corpora, which can be used to compute averages of their polynomial functions. The Gaussian matrix theory  we consider, and potential perturbations thereof one might consider in the future, are the theoretical analogs of the path integrals for inflation considered in cosmology. This natural science perspective on distributional semantics based on matrix models offers an interesting complement to the artificial intelligence perspectives which drive much research on distributional semantics. Beyond the question of quantifying Gaussianity, our investigations of   linguistic matrix data in this paper are guided by the intriguing interfaces between the two perspectives of natural science and artificial intelligence.

Going back to the first goal of this paper, we find that low order permutation invariant polynomials, and specifically the 13-parameter Gaussian permutation invariant matrix models, are indeed 
the right objects to detect strong evidence of Gaussianity. While the best theory/expt ratios achieved by the 5-parameter model are near $60 \%$, the best ratios now are near $ 99\% $ and indeed for a number of cubic and quartic observables, these ratios are above  $ 90 \% $. The lowest  ratio is $ 16 \%$, so that the Gaussian model still predicts the right order of magnitude of the expectation value 
even in the worst case. In all the experiments studied, we find that the linear and quadratic expectation values lead to  theoretical parameters $ \mu, \Lambda$ consistent with the convergence criteria. 

Since the comparison of experiment with theory in the above discussion has only used, for each observable, the experimental averages of the observables $ \cO ( M ) $ over all the words, it is oblivious to the detailed distribution of the observable over the set of words used to calculate the average. This distribution  has a standard deviation $ (\delta \cO ( M ))_{EXPT}  $. As a further test of Gaussianity, we can use the standard deviations of the linear and quadratic observables from the data, to determine 
perturbed theoretical parameters $ \mu + \delta \mu , \Lambda + \delta \Lambda $, and then use the 
theoretical equations to determine theoretical predictions $ (\delta \cO (M) )_{THEO} $ for the standard deviations of the higher order observables. We find that the theory/experiment ratios for the standard deviations range over $ 26 \% $ to $ 95 \% $. This looks to be a very good success rate, which we confirm by  comparing to a simple random walk model for the standard deviations.  In our prediction of the standard deviations, we are using for the same dataset, a range of possible values of the couplings in the Gaussian matrix model, effectively a Gaussian model with a distribution of couplings. The success of 
these predictions of the standard deviations is our second main result.

Our tables of theory/experiment ratios for $ \langle \cO ( M ) \rangle $ and $ \delta \cO ( M )  $ show that some pairs of observables have distinctly similar characteristics whether we are looking at  expectation values or standard deviations. Each observable can also be used to 
rank the words in the dataset, starting from the word with the lowest $ \cO ( M ) $ to the one with the highest. Since ranked lists of words form a standard tool in distributional semantics, it is natural to ask whether observables which have very similar matrix model characteristics also produce similar ranked lists. We find evidence for a positive answer.

The plan of the paper is as follows. 
Section \ref{sec:techintro} is a technical introduction describing the range of experimental data we will be analysing. Section \ref{sec:linsys} gives the system of equations for the theoretical expectation values of  the two linear and  eleven quadratic observables as a function of the theoretical parameters $ \mu  , \Lambda$. For each experiment, the thirteen experimental expectation values are matched with the theoretical ones, to determine the appropriate $  \mu , \Lambda  $ for each experiment. 
Section \ref{sec:GenDim} gives the ratios of experimental to theoretical expectation values for 
cubic and quartic observables.  Section \ref{sec:typicality} explains the motivations from 
possible applications in distributional semantics for our investigations of typicality. It then proceeds to explain the experiment/theory comparisons and presents the results. 
Section \ref{sec:rankings} provides evidence showing that observables with similar matrix model characteristics, in terms of expectation values and dispersions, produce similar ranked lists. 
The comparison of ranked lists is done with the Spearman $\rho$   characteristic  as well as two-dimensional rank correlation  plots. We conclude with a discussion of our results and future directions. 

Appendix  \ref{App:theorcubquart} lists the equations for  the expectation values of 
cubic and quartic observables  in terms of the theoretical parameters $ \mu , \Lambda $. 
A number of these equations  are reproduced, in one or two instances with typos corrected, 
 from \cite{PIGMM} and there are four new observables
which are computed by the same methods explained there.

\vskip.5cm

\section{ Experiments and the 13 theoretical Parameters   }\label{sec:techintro}

In this paper, we will be using the matrices for adjectives and verbs that were constructed for the paper \cite{LMT}. The detailed algorithm is explained there. The matrices are of size $ D \times D$, where 
$D$ ranges in steps of $100$ from $300$ to $ 2000$. 

As explained in \cite{LMT} the counting of linearly independent permutation invariant polynomials (observables)  of a fixed degree is equivalently given by the counting  of directed graphs. The nodes correspond to the indices, the matrix $M_{ij}$ correspond to an edge going from node $i$ to node $j$. The graphs corresponding to the  $11$ quadratic polynomials are given in Appendix B  of \cite{LMT}. One minor technical point : in this paper, we find it convenient to associate unrestricted sums to graphs, e.g. 
for a graph having two edges from one node to another, we associate $ \sum_{ i , j } M_{ij}^2$, and not 
$ \sum_{ i \ne j } M_{ ij}^2 $ as in \cite{LMT}.

There are $52$ cubic observables/graphs and $296$ quartic ones. In \cite{PIGMM} the thirteen parameter model was solved. The computation of expectation values was given for a selection of four cubic and two quartic observables. In this paper, we have developed the theoretical formulae for an additional four observables.  The graphs corresponding to the  set of  ten cubic/quartic observables under consideration 
in this paper are given in Appendix \ref{App:Graphdiagram}. The theoretical equations for the ten expectation values are given in Appendix \ref{App:theorcubquart}.

\subsection{ The system of equations  for the 13-parameters  }\label{sec:linsys}

The strategy for comparison of experiments with data we use here is exactly as in \cite{LMT}, with 
the only difference now being that we have the full $13$-dimensional parameter space of 
permutation invariant Gaussian matrix models.

From \cite{PIGMM}, 
the two  equations expressing expectation values of linear permutation invariant functions of $M$, in terms of the $ \mu ,\Lambda $ parameters of  the Gaussian model. 
\begin{align}
     \sum_{i} \langle M_{ii} \rangle &= \widetilde{\mu}_{1} + \sqrt{(D-1)}\widetilde{\mu}_{2}.
\end{align}

\begin{align}
     \sum_{i,j} \langle M_{ij} \rangle &= D\widetilde{\mu}_{1}.
\end{align}
where 
\bea 
&& \widetilde{\mu}_{1} = ( ( \Lambda_{ V_0} )^{-1} )_{ 11} \mu_1 + ( ( \Lambda_{ V_0} )^{-1} )_{ 12} \mu_2 \cr
&&  \widetilde{\mu}_{2} = ( ( \Lambda_{ V_0} )^{-1} )_{ 21} \mu_1 + ( ( \Lambda_{ V_0} )^{-1} )_{ 22} \mu_2
\eea

The  eleven equations \cite{PIGMM} expressing expectation values of quadratic 
 permutation invariant functions of $M$, in terms of
the $ \mu ,\Lambda $ parameters of  the Gaussian model are as follows. 

\begin{align}
  \sum_{i,j} \langle M_{ij} M_{ij} \rangle &= \widetilde{\mu}_{1}^{2} + \widetilde{\mu}_{2}^{2} + (\Lambda_{V_0}^{-1})_{11} + (\Lambda_{V_0}^{-1})_{22} + (D-1)(\Lambda_{H}^{-1})_{22} + (D-1)(\Lambda_{H}^{-1})_{33} \cr 
    &+ (D-1)(\Lambda_{H}^{-1})_{11} + \frac{D(D-3)}{2}(\Lambda_{V_2})^{-1} + \frac{(D-1)(D-2)}{2} (\Lambda_{V_3})^{-1}.
\end{align}

\begin{align}
     \sum_{i,j} \langle M_{ij}M_{ji} \rangle &= (\Lambda_{V_2})^{-1}\frac{D(D-3)}{2} - (\Lambda_{V_3})^{-1}\frac{(D-1)(D-2)}{2} + 2(D-1)(\Lambda_{H}^{-1})_{12} + (D-1)(\Lambda_{H}^{-1})_{33} \cr 
      &+ (\Lambda_{V_0}^{-1})_{11} + (\Lambda_{V_0}^{-1})_{22} + \widetilde{\mu}_{1}^{2} + \widetilde{\mu}_{2}^{2}.
\end{align}

\begin{align}
     \sum_{i,j} \langle M_{ii}M_{ij} \rangle &= (\Lambda_{V_0}^{-1})_{11} + \sqrt{(D-1)}(\Lambda_{V_0}^{-1})_{12} + (D-1)(\Lambda_{V_H}^{-1})_{12} + (D-1)(\Lambda_{V_H}^{-1})_{22} \cr 
      &+ (D-1)\sqrt{(D-2)}(\Lambda_{V_H}^{-1})_{23} + \widetilde{\mu}_{1}^{2} + \widetilde{\mu}_{1}\widetilde{\mu}_{2} \sqrt{(D-1)}.
\end{align}

\begin{align}
    \sum_{i,j} \langle M_{ii} M_{ji} \rangle &= (\Lambda_{V_0}^{-1})_{11} + \sqrt{(D-1)}(\Lambda_{V_0}^{-1})_{12}  + (D-1)(\Lambda_{V_H}^{-1})_{11} + (D-1)(\Lambda_{V_H}^{-1})_{12} \cr  &+ (D-1)\sqrt{(D-2)}(\Lambda_{V_H}^{-1})_{13} +
     \widetilde{\mu}_{1}^2 +  \widetilde{\mu}_{1}  \widetilde{\mu}_{2}\sqrt{(D-1)}.  
\end{align}

\begin{align}
    \sum_{i,j,k} \langle M_{ij} M_{ik} \rangle &= D (\Lambda_{V_0}^{-1})_{11} + D(D-1) (\Lambda_{V_H}^{-1})_{22} + D\widetilde{\mu}_{1}^2.
\end{align}

\begin{align}
     \sum_{i,j,k} \langle M_{ij} M_{kj} \rangle &= D (\Lambda_{V_0}^{-1})_{11} + D(D-1) (\Lambda_{V_H}^{-1})_{11} + D\widetilde{\mu}_{1}^2.
\end{align}

\begin{align}
    \sum_{i,j,k} \langle M_{ij} M_{jk} \rangle &= D (\Lambda_{V_0}^{-1})_{11} + D(D-1) (\Lambda_{V_H}^{-1})_{12} + D\widetilde{\mu}_{1}^2.
\end{align}

\begin{align}
    \sum_{i,j,k,l} \langle M_{ij} M_{kl} \rangle &= D^2 (\Lambda_{V_0}^{-1})_{11} + D^2\widetilde{\mu}_{1}^2.
\end{align}

\begin{align}
    \sum_{i} \langle M_{ii}^2 \rangle &= \frac{(\Lambda_{V_0}^{-1})_{11}}{D} + \frac{(D-1)}{D}(\Lambda_{V_0}^{-1})_{22} + 2\frac{\sqrt{(D-1)}}{D}(\Lambda_{V_0}^{-1})_{12} + \frac{(D-1)}{D}(\Lambda_{V_H}^{-1})_{11} +\frac{(D-1)}{D}(\Lambda_{V_H}^{-1})_{22} \cr 
    &+ \frac{(D-1)}{D}(D-2)(\Lambda_{V_H}^{-1})_{33} + 2\frac{(D-1)}{D}(\Lambda_{V_H}^{-1})_{12} + 2\frac{(D-1)}{D}\sqrt{(D-2)}(\Lambda_{V_H}^{-1})_{13} \cr 
    &+2\frac{(D-1)}{D}\sqrt{(D-2)}(\Lambda_{V_H}^{-1})_{23}
    + \frac{\widetilde{\mu}_{1}^2}{D} + 2\frac{\widetilde{\mu}_{1}\widetilde{\mu}_{2}}{D}\sqrt{(D-1)} + \widetilde{\mu}_{2}^2 \frac{(D-1)}{D}.
\end{align}

\begin{align}
    \sum_{i,j} \langle M_{ii}M_{jj} \rangle &= (\Lambda_{V_0}^{-1})_{11} + (D-1)(\Lambda_{V_0}^{-1})_{22} + 2\sqrt{(D-1)}(\Lambda_{V_0}^{-1})_{12} + \widetilde{\mu}_{1}^2 + 2\widetilde{\mu}_{1}\widetilde{\mu}_{2}\sqrt{(D-1)} \cr 
     &+  \widetilde{\mu}_{2}^2 (D-1).
\end{align}

\begin{align}
    \sum_{i,j,k} \langle M_{ii}M_{jk} \rangle &= D(\Lambda_{V_0}^{-1})_{11} + D\sqrt{(D-1)}(\Lambda_{V_0}^{-1})_{12} + D\widetilde{\mu}_{1}^2 + \widetilde{\mu}_{1}\widetilde{\mu}_{2}D\sqrt{(D-1)}.
\end{align}

\subsection{ Parameter values for adjectives at $ D=2000 $}

To 3 significant figures, the parameter values for $ D = 2000$ are given below. 

\begin{center}
\begin{tabular}{ | c | c | } 
\hline
Parameter & Value \\[0.3mm]
\hline
$\widetilde{\mu_{1}} $ & $\phantom{a}4.84 \times 10^{-1}$ \\[0.3mm]
\hline
$\widetilde{\mu_{2}} $ & $1.01$ \\[0.3mm]
\hline
$(\Lambda_{V_0}^{-1})_{11}$ & $4.00 \times 10^{-2}$ \\[0.3mm]
\hline
$(\Lambda_{V_0}^{-1})_{12}$ & $5.10 \times 10^{-2}$ \\[0.3mm]
\hline
$(\Lambda_{V_0}^{-1})_{22}$ & $2.49 \times 10^{-1} $\\[0.3mm]
\hline
$(\Lambda_{H}^{-1})_{11}$ & $ 1.45 \times 10^{-2}$\\[0.3mm]
\hline
$(\Lambda_{H}^{-1})_{12}$ & $1.02 \times 10^{-4}$  \\[0.3mm]
\hline
$(\Lambda_{H}^{-1})_{13}$ & $2.28 \times 10^{-4}$ \\[0.3mm]
\hline
$(\Lambda_{H}^{-1})_{22}$ & $ 2.91 \times 10^{-4}$ \\[0.3mm]
\hline
$(\Lambda_{H}^{-1})_{23}$ &  $ 1.22 \times 10^{-4}$  \\[0.3mm]
\hline
$(\Lambda_{H}^{-1})_{33}$ & $ 7.27 \times 10^{-4} $\\[0.3mm]
\hline
$(\Lambda_{V_{2}}^{-1})$ &  $ 2.49 \times 10^{-4}$\\[0.3mm]
\hline
$(\Lambda_{V_{3}}^{-1})$ & $2.41 \times 10^{-4}$\\[0.3mm]
\hline
\end{tabular}
\end{center}

 The  values of the determinants of the coupling matrices for each irreducible representation of $S_D$, calculated by entering the experimental linear and quadratic expectation values into the system of 
 equations in Section \ref{sec:linsys}, are  (to 3 significant figures) 
$$\text{Det}(\Lambda_{V_0}) = 1.36 \times 10^{2}$$
$$\text{Det}(\Lambda_{V_H}) = 3.54 \times 10^{8}$$
$$\Lambda_{V_2} = 4.02 \times 10^{3}$$
$$\Lambda_{V_3} = 4.14 \times 10^{3}$$
Since these are all positive, the criteria are satisfied.  This is evidence for the Gaussian ansatz.

\subsection{Parameter values for verbs at $ D = 2000$ }

The parameters of the model (to three significant figures) calculated  are

\begin{center}
\begin{tabular}{ | c | c | } 
\hline
Parameter & Value \\[0.3mm]
\hline
$\tilde{\mu_{1}} $ & $4.29 \times 10^{-1}$ \\[0.3mm]
\hline
$\tilde{\mu_{2}} $ & $1.06$ \\[0.3mm]
\hline
$(\Lambda_{V_0}^{-1})_{11}$ & $5.52 \times 10^{-2}$ \\[0.3mm]
\hline
$(\Lambda_{V_0}^{-1})_{12}$ & $4.68 \times 10^{-2}$ \\[0.3mm]
\hline
$(\Lambda_{V_0}^{-1})_{22}$  & $2.86 \times 10^{-1}$ \\[0.3mm]
\hline
$(\Lambda_{H}^{-1})_{11}$ & $1.84 \times 10^{-2}$\\[0.3mm]
\hline
$(\Lambda_{H}^{-1})_{12}$ & $7.65 \times 10^{-5}$ \\[0.3mm]
\hline
$(\Lambda_{H}^{-1})_{13}$ & $2.85 \times 10^{-4}$ \\[0.3mm]
\hline
$(\Lambda_{H}^{-1})_{22}$ & $2.30 \times 10^{-4}$\\[0.3mm]
\hline
$(\Lambda_{H}^{-1})_{23}$ & $9.34 \times 10^{-5}$\\[0.3mm]
\hline
$(\Lambda_{H}^{-1})_{33}$ & $8.62 \times 10^{-4}$ \\[0.3mm]
\hline
$(\Lambda_{V_{2}}^{-1})$ & $3.08 \times 10^{-4}$ \\[0.3mm]
\hline
$(\Lambda_{V_{3}}^{-1})$ & $3.00 \times 10^{-4}$\\[0.3mm]
\hline
\end{tabular}
\end{center}

Convergence criteria for verbs are also satisfied:

$$\text{Det}(\Lambda_{V_0}) = 73.6$$
$$\text{Det}(\Lambda_{V_H}) = 2.89 \times 10^{8}$$
$$\Lambda_{V_2} = 3.25 \times 10^{3}$$
$$\Lambda_{V_3} = 3.33 \times 10^{3}$$

\section{ Theory/Expt comparisons for expectation values of observables : evidence for Gaussianity  } 
\label{sec:expect}

In this section, we  describe the comparisons for expectation values of cubic and quartic observables. 
We find significant agreements at very high levels of accuracy, in the range $90-99 \% $ 
for a number of observables. This is to be compared with the $57 \%$ accuracies that were achieved as the optimum ratios with the 5-parameter model \cite{LMT}. 

The lowest theory/expt ratio  with the $13$-parameter model is at $16 \%$. So we have the right order of magnitude even in this worst case.  

There are regularities in the nature of high ratio versus low ratio observables, in terms of simple 
characteristics of the observable-graph, notably the number of nodes.  The very high Gaussianities, reflected in ratios $ \langle \cO ( M ) \rangle_{THEO} / \langle \cO ( M ) \rangle_{EXPT} $  close to $1$ occur for graphs with four or more nodes.  The number of nodes corresponds to the number of indices being summed, hence also to a $D$-scaling of the number of terms in the defining sum. 

In detail, the results for the  Cubic and Quartic ratios for 13 parameter model are given below. 
The first table is for the matrices associated with adjectives, while the second is for verbs. 

\noindent 
{\bf Adjectives  at D = 2000 :}

\begin{center} 
\begin{tabular}{ | c | c | c | c | c | }
\hline
Graph & Expectation value & Theoretical val. & Experimental val. & Ratio\\ 
\hline
1 & $\sum_{i} \langle (M_{ii})^3 \rangle$ & $1.44 \times 10^{-1}$ & $2.52 \times 10^{-1}$ & 0.57\\ 
\hline
2 & $\sum_{i,j} \langle (M_{ij})^3 \rangle$ & $8.43 \times 10^{-1}$ & 3.65 & 0.23 \\ 
\hline
3 & $\sum_{i,j,k} \langle M_{ij}M_{jk}M_{ki} \rangle$ & 1.68 & 10.6 & 0.16\\
\hline
4 & $\sum_{i,j,k} \langle M_{ij}M_{jj}M_{jk} \rangle$ & 53.8 & 80.1 & 0.67\\
\hline
5 & $\sum_{i,j,k,l} \langle M_{ij}M_{kk}M_{ll} \rangle$ & $2.94 \times 10^{6}$ & $3.03 \times 10^{6}$ & 0.97\\ 
\hline
6 & $\sum_{i,j,k,l} \langle M_{ij}M_{jk}M_{ll} \rangle$ & $4.83 \times 10^{4}$ & $5.04 \times 10^{4}$ & 0.96\\ 
\hline
7 & $\sum_{i,j,k,l,m} \langle M_{ij}M_{kl}M_{mm} \rangle$ & $5.93 \times 10^{7}$ & $6.01 \times 10^{7}$ & 0.99 \\ 
\hline
8 & $\sum_{i,j,k,l,m,n} \langle M_{ij}M_{kl}M_{mn} \rangle$ & $1.38 \times 10^{9}$ & $1.40  \times 10^{9}$ & 0.98 \\ 
\hline
9 & $\sum_{i_{1}...i_{7}}\langle M_{i_{1}i_{2}}M_{i_{3}i_{4}}M_{i_{5}i_{6}}M_{i_{7}i_{7}} \rangle$ & $7.83 \times 10^{10}$  & $8.14 \times 10^{10}$  & 0.96  \\
\hline
10 & $\sum_{i_{1}...i_{8}}\langle M_{i_{1}i_{2}}M_{i_{3}i_{4}}M_{i_{5}i_{6}}M_{i_{7}i_{8}} \rangle$ & $1.86 \times 10^{12}$ & $1.96 \times 10^{12}$ & 0.95 \\ 
\hline
\end{tabular}
\end{center}

\noindent 
{\bf Verbs at D = 2000 }

\begin{center}
\begin{tabular}{ | c | c | c | c | c | }
\hline
Graph & Expectation value & Theoretical val. & Experimental val. & Ratio\\ 
\hline
1 & $ \sum_{i} \langle (M_{ii})^3 \rangle $ & $1.76 \times 10^{-1}$ & $3.22 \times 10^{-1}$ & 0.55\\ 
\hline
2 & $\sum_{i,j} \langle (M_{ij})^3 \rangle $ & $9.36 \times 10^{-1}$ & 4.26 & 0.22\\ 
\hline
3 & $ \sum_{i,j,k}\langle M_{ij}M_{jk}M_{ki} \rangle$ & 1.62 & 9.98 & 0.16\\ 
\hline
4 & $ \sum_{i,j,k}\langle M_{ij}M_{jj}M_{jk} \rangle$ & 51.2 & 73.7 & 0.70\\ 
\hline
5 & $\sum_{i,j,k,l} \langle M_{ij}M_{kk}M_{ll} \rangle$ & $2.87 \times 10^{6}$  & $2.92 \times 10^{6}$  & 0.99 \\ 
\hline
6 & $\sum_{i,j,k,l} \langle M_{ij}M_{jk}M_{ll} \rangle$ & $4.12 \times 10^{4}$  & $4.32 \times 10^{4}$  & 0.95 \\ 
\hline
7 & $\sum_{i,j,k,l,m} \langle M_{ij}M_{kl}M_{mm} \rangle$ & $5.32 \times 10^{7}$ & $5.35 \times 10^{7}$ & 0.99  \\ 
\hline
8 & $\sum_{i,j,k,l,m,n} \langle M_{ij}M_{kl}M_{mn} \rangle $ & $1.20 \times 10^{9}$ & $1.26 \times 10^{9}$ & 0.95 \\ 
\hline
9 & $\sum_{i_{1}...i_{7}}\langle M_{i_{1}i_{2}}M_{i_{3}i_{4}}M_{i_{5}i_{6}}M_{i_{7}i_{7}} \rangle$ & $6.97 \times 10^{10}$ & $7.27 \times 10^{10}$ & 0.96  \\
\hline
10 & $\sum_{i_{1}...i_{8}}\langle M_{i_{1}i_{2}}M_{i_{3}i_{4}}M_{i_{5}i_{6}}M_{i_{7}i_{8}} \rangle$ & $1.66 \times 10^{12}$ & $1.85 \times 10^{12}$ & 0.90 \\ 
\hline
\end{tabular}
\end{center}

\section{ Results as a function of dimension }\label{sec:GenDim} 

\begin{figure}[h!]
\centering
\includegraphics[width=12cm,height=9cm]{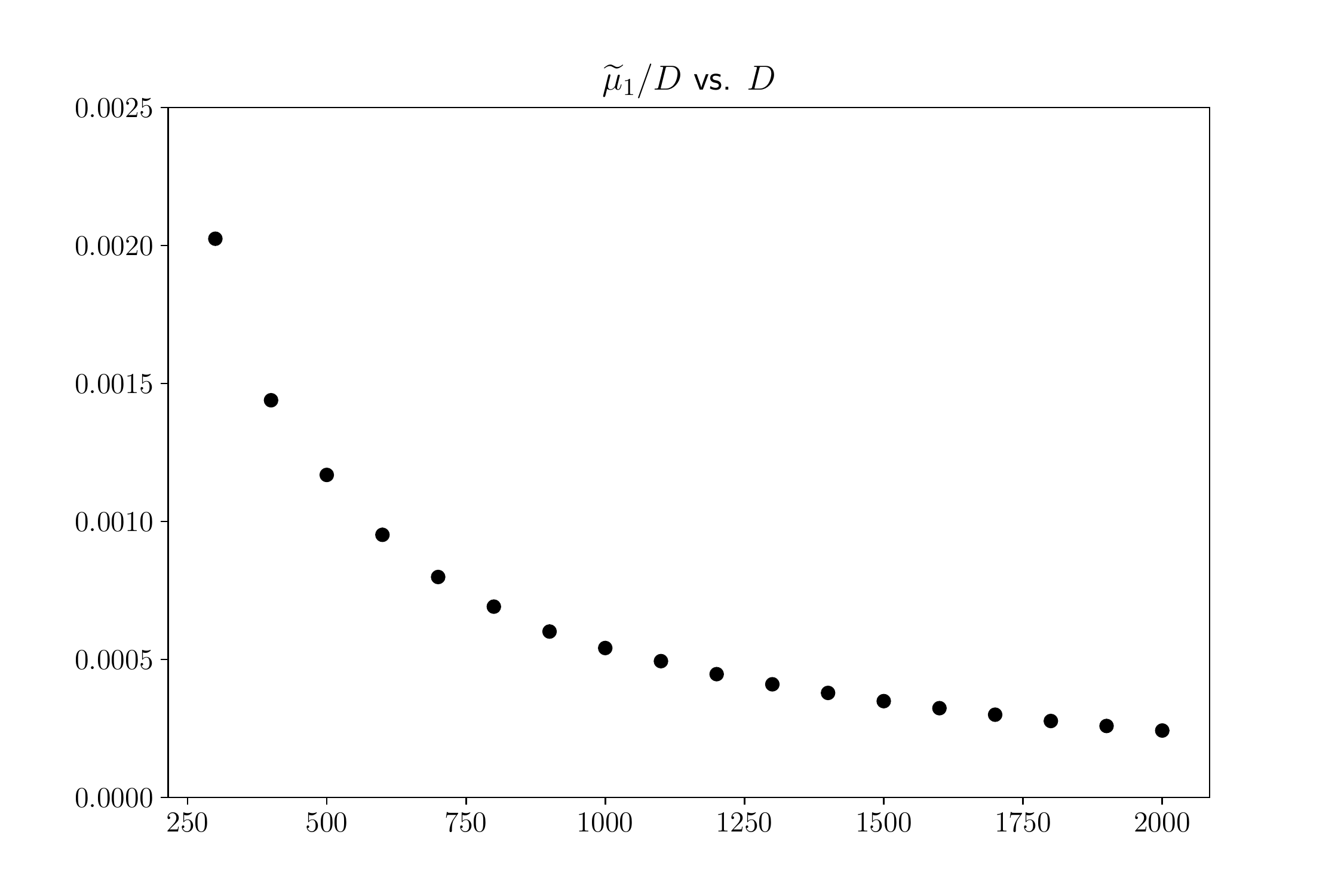}
\caption{Parameter $\frac{\widetilde{\mu}_{1}}{D}$ value vs. dimension $D$}
\label{fig:Param_mu_1_div_D}
\end{figure}

\begin{figure}[h!]
\centering
\includegraphics[width=12cm,height=9cm]{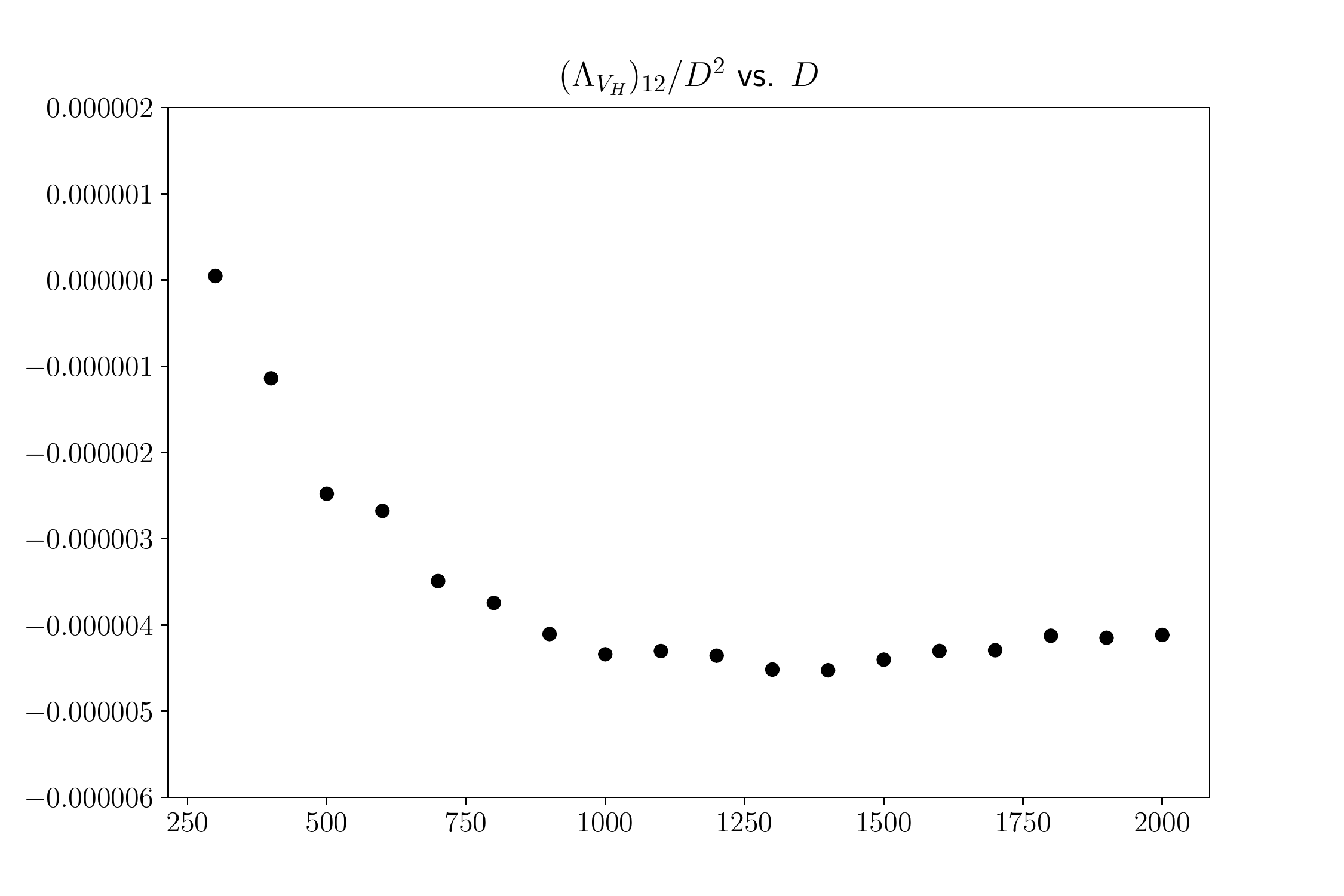}
\caption{Parameter $\frac{(\Lambda_{V_{H}})_{12}}{D^{2}}$ value vs. dimension $D$}
\label{fig:Param_VH_12_div_Dsquared}
\end{figure}

 Upon further testing, the convergence criteria for all dimensions of both verb and adjective data sets were confirmed to be satisfied. The explicit criteria calculation values for the dimensions 700 and 1300 are provided below.
The parameters can also be cast a function of $D$ and plotted to evaluate the dependence. Included here are example plots for two selected parameters, detailing their value for dimensions ranging from $300$ to $2000$ (see figures \ref{fig:Param_mu_1_div_D} and \ref{fig:Param_VH_12_div_Dsquared}).

We tend to see an onset of simple scaling behaviours at around $ D = 700$, hence we also present the calculations of the theory/expt ratios for  $ D = 700$ and $ D = 1300$.

\newpage

{\bf Adjectives at D = 700}

\begin{center}
\begin{tabular}{ | c | c | c | c | c | }
\hline
Graph & Expectation value & Theoretical val. & Experimental val. & Ratio\\ 
\hline
1 & $\sum_{i} \langle (M_{ii})^3 \rangle$ & 2.11 & 3.24 & 0.65\\ 
\hline
2 & $\sum_{i,j} \langle (M_{ij})^3 \rangle$ & 5.89 & 14.7 & 0.40 \\ 
\hline
3 & $\sum_{i,j,k} \langle M_{ij}M_{jk}M_{ki} \rangle$ & 7.58 & 23.3 & 0.33\\
\hline
4 & $\sum_{i,j,k} \langle M_{ij}M_{jj}M_{jk} \rangle$ & 96.1 & $1.28 \times 10^{2}$ & 0.75\\
\hline
5 & $\sum_{i,j,k,l} \langle M_{ij}M_{kk}M_{ll} \rangle$ & $1.90 \times 10^{6}$ & $1.97 \times 10^{6}$ & 0.96\\ 
\hline
6 & $\sum_{i,j,k,l} \langle M_{ij}M_{jk}M_{ll} \rangle$ & $2.63 \times 10^{4}$ & $2.64 \times 10^{4}$ & 0.998\\ 
\hline
7 & $\sum_{i,j,k,l,m} \langle M_{ij}M_{kl}M_{mm} \rangle$ & $1.24 \times 10^{7}$ & $1.27 \times 10^{7}$ & 0.98 \\ 
\hline
8 & $\sum_{i,j,k,l,m,n} \langle M_{ij}M_{kl}M_{mn} \rangle$ & $9.47 \times 10^{7}$ & $9.75 \times 10^{7}$ & 0.97 \\ 
\hline
9 & $\sum_{i_{1}...i_{7}}\langle M_{i_{1}i_{2}}M_{i_{3}i_{4}}M_{i_{5}i_{6}}M_{i_{7}i_{7}} \rangle$ & $6.74 \times 10^{9}$  & $7.25 \times 10 ^{9}$  & 0.93  \\
\hline
10 & $\sum_{i_{1}...i_{8}}\langle M_{i_{1}i_{2}}M_{i_{3}i_{4}}M_{i_{5}i_{6}}M_{i_{7}i_{8}} \rangle$ & $5.33 \times 10^{10}$ & $5.84 \times 10^{10}$ & 0.91 \\ 
\hline
\end{tabular}
\end{center}

Convergence criteria:
$$\text{Det}(\Lambda_{V_0}) = 18.0$$
$$\text{Det}(\Lambda_{V_H}) = 1.15 \times 10^{6}$$
$$\Lambda_{V_2} = 2.21 \times 10^{2}$$
$$\Lambda_{V_3} = 2.77 \times 10^{2}$$

{\bf Adjectives at D = 1300}

\begin{center}
\begin{tabular}{ | c | c | c | c | c | }
\hline
Graph & Expectation value & Theoretical val. & Experimental val. & Ratio\\ 
\hline
1 & $\sum_{i} \langle (M_{ii})^3 \rangle$ & $4.53 \times 10^{-1}$ & $7.39 \times 10^{-1}$ & 0.61\\ 
\hline
2 & $\sum_{i,j} \langle (M_{ij})^3 \rangle$ & 1.89 & 6.54 & 0.29 \\ 
\hline
3 & $\sum_{i,j,k} \langle M_{ij}M_{jk}M_{ki} \rangle$ & 3.25 & 15.1 & 0.21\\
\hline
4 & $\sum_{i,j,k} \langle M_{ij}M_{jj}M_{jk} \rangle$ & 72.7 & $1.03 \times 10^{2}$ & 0.71\\
\hline
5 & $\sum_{i,j,k,l} \langle M_{ij}M_{kk}M_{ll} \rangle$ & $2.75 \times 10^{6}$ & $2.86 \times 10^{6}$  & 0.96\\ 
\hline
6  & $\sum_{i,j,k,l} \langle M_{ij}M_{jk}M_{ll} \rangle$ & $4.27 \times 10^{4}$ & $4.48 \times 10^{4}$ & 0.95\\ 
\hline
7 & $\sum_{i,j,k,l,m} \langle M_{ij}M_{kl}M_{mm} \rangle$ & $3.57 \times 10^{7}$ & $3.65 \times 10^{7}$ & 0.98 \\ 
\hline
8 & $\sum_{i,j,k,l,m,n} \langle M_{ij}M_{kl}M_{mn} \rangle$ & $5.30 \times 10^{8}$ & $5.46 \times 10^{8}$ & 0.97 \\ 
\hline
9 & $\sum_{i_{1}...i_{7}}\langle M_{i_{1}i_{2}}M_{i_{3}i_{4}}M_{i_{5}i_{6}}M_{i_{7}i_{7}} \rangle$ & $3.49 \times 10^{10}$  & $3.74 \times 10^{10}$ & 0.93  \\
\hline
10 & $\sum_{i_{1}...i_{8}}\langle M_{i_{1}i_{2}}M_{i_{3}i_{4}}M_{i_{5}i_{6}}M_{i_{7}i_{8}} \rangle$ & $5.32 \times 10^{11}$ & $5.79 \times 10^{11}$ & 0.92 \\ 
\hline
\end{tabular}
\end{center}

Convergence criteria:
$$\text{Det}(\Lambda_{V_0}) = 50.7$$
$$\text{Det}(\Lambda_{V_H}) = 3.44 \times 10^{7}$$
$$\Lambda_{V_2} = 1.36 \times 10^{3}$$
$$\Lambda_{V_3} = 1.40 \times 10^{3}$$

\noindent 
{\bf Verbs at  D = 700  } 

\begin{center}
\begin{tabular}{ | c | c | c | c | c | }
\hline
Graph & Expectation value & Theoretical val. & Experimental val. & Ratio\\ 
\hline
1 & $\sum_{i} \langle (M_{ii})^3 \rangle$ & 2.52 & 4.05 & 0.62\\ 
\hline
2 & $\sum_{i,j} \langle (M_{ij})^3 \rangle$ & 8.06 & 18.7 & 0.43 \\ 
\hline
3 & $\sum_{i,j,k} \langle M_{ij}M_{jk}M_{ki} \rangle$ & 7.61 & 16.1 & 0.47\\
\hline
4 & $\sum_{i,j,k} \langle M_{ij}M_{jj}M_{jk} \rangle$ & $1.12 \times 10^{2}$ & $1.46 \times 10^{2}$ & 0.77\\
\hline
5 & $\sum_{i,j,k,l} \langle M_{ij}M_{kk}M_{ll} \rangle$ & $2.24 \times 10^{6}$ & $2.32 \times 10^{6}$ & 0.97\\ 
\hline
6  & $\sum_{i,j,k,l} \langle M_{ij}M_{jk}M_{ll} \rangle$ & $2.86 \times 10^{4}$ & $2.95 \times 10^{4}$ & 0.97\\ 
\hline
7 & $\sum_{i,j,k,l,m} \langle M_{ij}M_{kl}M_{mm} \rangle$ & $1.59 \times 10^{7}$ & $1.62 \times 10^{7}$ & 0.98 \\ 
\hline
8 & $\sum_{i,j,k,l,m,n} \langle M_{ij}M_{kl}M_{mn} \rangle$ & $1.32 \times 10^{8}$ & $1.37 \times 10^{8}$ & 0.97 \\ 
\hline
9 & $\sum_{i_{1}...i_{7}}\langle M_{i_{1}i_{2}}M_{i_{3}i_{4}}M_{i_{5}i_{6}}M_{i_{7}i_{7}} \rangle$ & $1.00 \times 10^{10}$  & $1.07 \times 10^{10}$ & 0.94  \\
\hline
10 & $\sum_{i_{1}...i_{8}}\langle M_{i_{1}i_{2}}M_{i_{3}i_{4}}M_{i_{5}i_{6}}M_{i_{7}i_{8}} \rangle$ & $8.65 \times 10^{10}$ & $ 9.30 \times 10^{10}$  & 0.93 \\ 
\hline
\end{tabular}
\end{center}

Convergence criteria:
$$\text{Det}(\Lambda_{V_0}) = 11.1$$
$$\text{Det}(\Lambda_{V_H}) = 7.81 \times 10^{5}$$
$$\Lambda_{V_2} = 2.00 \times 10^{2}$$
$$\Lambda_{V_3} = 2.03 \times 10^{2}$$

\noindent 
{\bf Verbs at  D  = 1300  } 

\begin{center}
\begin{tabular}{ | c | c | c | c | c | }
\hline
Graph & Expectation value & Theoretical val. & Experimental val. & Ratio\\ 
\hline
1 & $\sum_{i} \langle (M_{ii})^3 \rangle$ & $5.48 \times 10^{-1}$ & $9.34 \times 10^{-1}$ & 0.59\\ 
\hline
2 & $\sum_{i,j} \langle (M_{ij})^3 \rangle$ & 2.30 & 7.77 & 0.30 \\ 
\hline
3 & $\sum_{i,j,k} \langle M_{ij}M_{jk}M_{ki} \rangle$ & 3.27 & 14.0 & 0.23\\
\hline
4 & $\sum_{i,j,k} \langle M_{ij}M_{jj}M_{jk} \rangle$ & 76.2 & $1.04 \times 10^{2}$ & 0.73\\
\hline
5 & $\sum_{i,j,k,l} \langle M_{ij}M_{kk}M_{ll} \rangle$ & $2.86 \times 10^{6}$ & $2.91 \times 10^{6}$ & 0.98\\ 
\hline
6  & $\sum_{i,j,k,l} \langle M_{ij}M_{jk}M_{ll} \rangle$ & $4.04 \times 10^{4}$ & $4.24 \times 10 ^{4}$ & 0.95\\ 
\hline
7 & $\sum_{i,j,k,l,m} \langle M_{ij}M_{kl}M_{mm} \rangle$ & $3.64 \times 10^{7}$ & $3.67 \times 10^{7}$ & 0.99 \\ 
\hline
8 & $\sum_{i,j,k,l,m,n} \langle M_{ij}M_{kl}M_{mn} \rangle$ & $5.58 \times 10^{8}$ & $5.77 \times 10^{8}$ & 0.97 \\ 
\hline
9 & $\sum_{i_{1}...i_{7}}\langle M_{i_{1}i_{2}}M_{i_{3}i_{4}}M_{i_{5}i_{6}}M_{i_{7}i_{7}} \rangle$ & $3.69 \times 10^{10}$  & $3.84 \times 10^{10}$  & 0.96  \\
\hline
10 & $\sum_{i_{1}...i_{8}}\langle M_{i_{1}i_{2}}M_{i_{3}i_{4}}M_{i_{5}i_{6}}M_{i_{7}i_{8}} \rangle$ & $5.93 \times 10^{11}$ & $6.43 \times 10^{11}$  & 0.92 \\ 
\hline
\end{tabular}
\end{center}

Convergence criteria:
$$\text{Det}(\Lambda_{V_0}) = 28.9$$
$$\text{Det}(\Lambda_{V_H}) = 2.65 \times 10^{7}$$
$$\Lambda_{V_2} = 1.08 \times 10^{3}$$
$$\Lambda_{V_3} = 1.10 \times 10^{3}$$

\noindent 
\textbf{Remark} It is  worth mentioning that some of the cubic ratios, which are very low at $ D=2000$,  improve at $ D = 700$. The construction  of vectors for nouns and noun phrases, which is subsequently used to construct matrices for adjectives and verbs, relies on identifying sets of target nouns $t$  along with some context words $c$. There is a reasonable and well-defined prescription for dealing with the cases where target is equal to context word \cite{LMT}. However, these cases are perhaps more subtle. It is conceivable  that the low theory/experiment ratios for higher $D$ might be due to higher number of $ c=t$ cases. This can be investigated by repeating the experiments with datasets which have filtered out the $ c=t$ cases. We hope to return to this investigation in the future.

\section{ Typicality }\label{sec:typicality}

We have found that the postulate of Gaussianity allows the prediction, to a high degree of accuracy, 
of expectation values of a large number of cubic and quartic observables in type-driven compositional distributional semantics. These expectation values are calculated by taking averages over large numbers of large matrices, one for each adjective/verb.   A more detailed characterisation of the data in type-driven compositional distributional semantics gives, for each observable, a distribution of frequencies over a space of possible values of the observable. This can be visualized in terms of a histogram for each observable. The mean of the distribution is the expectation value but we may also look at the spread or variance of the observable.  We may ask for example whether these distributions become very narrow in the limit of large $D$.  The observables are sums of large numbers of matrix elements, these numbers being $ D^{ p}$ 
for some $p$, which is a characteristic of each observable. $p$ is in fact the number of indices in the sums, which is equal to the number of nodes in the graph. For example for the first graph/observable 
in the tables of Sections \ref{sec:expect} \ref{sec:GenDim}, we have $p=1$, while for the last observable we have $p=8$. When we calculate the observables, normalized by $ D^p$, and assume a simplistic model of random walks \cite{Wiki-RW}  where each term in the sum is a step, then we would have standard deviations of order $D^{p/2}$. This simplistic model suggests that the standard deviations of the $D^{p}$-normalised observables would behave like $ D^{ -p/2}$ and thus vanish at large $d$. This can also be argued as a consequence of the ``law of large numbers'' \cite{Wiki-LLN}. The qualitative expectation of a vanishing 
of the standard deviations in the limit of large numbers is indeed consistent with the standard deviations we find - so in this sense the distributions are consistent with typicality, in other words the distributions become peaked at large $D$. It is interesting, however, to ask if we can get a more precise prediction of the standard deviations observed in the permutation invariant observables using the permutation invariant Gaussian matrix models. A precise understanding of these standard deviations, or degrees of typicality for each observable, is motivated both by theoretical physics and the AI goals of 
distributional semantics.

In many physical systems with large numbers of degrees of freedom, considerations of  typicality are of fundamental interest. A common aspect in discussions of typicality is the statement that a large majority of the members of a large collection share some specified characteristic \cite{Frigg09}. 
A typicality characteristic of quantum states of composite systems, made of a physical system of interest with its environment, is explained as the origin of thermodynamic equilibrium states in quantum statistical thermodyanmics \cite{PSW0511,GLTZ0511,Frigg09}. Typicality has also been used in the context of the AdS/CFT correspondence as a proposal to account for the emergence of gravitational thermal states such as black holes, or closely related ``superstar geometries'' \cite{BBJS-0508,BCHLS0712}.

There are also practical motivations from computational linguistics for a systematic understanding of the typicality properties of the observables.
The construction of matrices of type-driven  distributional semantics, have 
applications to word and sentence similarity, disambiguation, and inference  tasks \cite{GS2015,KS2012,KSB2017}. In the similarity tasks, the goal is to decide how similar a pair of language units, such as words, phrases, sentences, and eventually  paragraphs and texts are to each other. Examples of sentence similarity  from the  three bands of HIGH, MED, and LOW similarity are the following pairs
\begin{quote}
 \texttt{(Project presented  problem, Report discussed  difficulties)}: HIGH \\
 \texttt{(Gentleman closed his eyes, man shot the door)}: MED  \\
 \texttt{(Project presented  problem, Gentleman closed his eyes)}: LOW
 \end{quote}
 Human judgements for these pairs  are collected, often using a crowd sourcing engine such as the Amazon Turk, and the degree of correlation between these judgements and the model measurements are computed. The degree of correlation  often used is the  Spearman's $\rho$, which is calculated between the two sets of values of average human judgements per pair of sentences, and the measurement of the model for the same pair, mainly via computing the cosine of the angle between the vectors of the sentences of the pair.  At the word level,  in a type-driven setting one builds matrices for adjectives and intransitive verbs (and cubes and hyper cubes  for transitive and ditransitive verbs) and computes the degree of correlation between the human annotations and the model similarity measures, see \cite{MC2015} for an adjective similarity task on the adjective subset of the MEN word similarity   dataset \cite{Brunetal2014} and \cite{BSJ2017} for a verb similarity  task on the VerbSim3500 verb similarity task \cite{Gerzetal2016}.  The inference task is slightly different, in that instead of a degree of similarity  in the unit interval one works with a Boolean value: 1 indicates that the first sentence entails the second one, as in the pair  \begin{quote}
 \texttt{(A cat danced, An animal moved)}
 \end{quote}
 and 0 says that it does not, as in 
 \begin{quote}
 \texttt{(A cat danced, The report presented a problem)}.
 \end{quote}
Here, asymmetric   measures,  such the Kullback–Leibler divergence,  are computed and compared with the Boolean measures.

An important problem in all of these tasks is to devise ways to efficiently construct the matrices for the large collection of words that have  functional types. Recall that these are the majority of the words of a language, ranging over adjectives, verbs, adverbs, wh-words, auxiliaries, and many more. The methods of constructing matrices are computationally expensive: one has to first parse the corpora of data to tag the words with their grammatical types, aka their part of speech or POS tags. This procedure will determine which words have atomic types and which ones are functional. Despite recent  advances in parsers via the use of neural network algorithms, these procedures are still erroneous  and given the large quantities of data that are needed to build the matrix, they will take long periods of  time to train. The question is whether we can supplement existing algorithms 
 for  producing the matrices by using universal  statistical characteristics  of the existing ones.
   It is conceivable that  the methods of linguistic matrix theory can be used to 
aid the construction. Imagine a sample of adjectives has been constructed. We would then determine the expectation values of some observables from the data of these matrices. Suppose then that the observable in question is a high typicality observable. Then if we wish to construct a new word matrix, we can devise algorithms which takes this predicted average as an input. 
This would require the development of algorithms which construct the matrices, but constrain 
their values for these high typicality observables to be very near the known averages.

\begin{figure}[ht!]
\centering
\includegraphics[width=14cm,height=11cm]{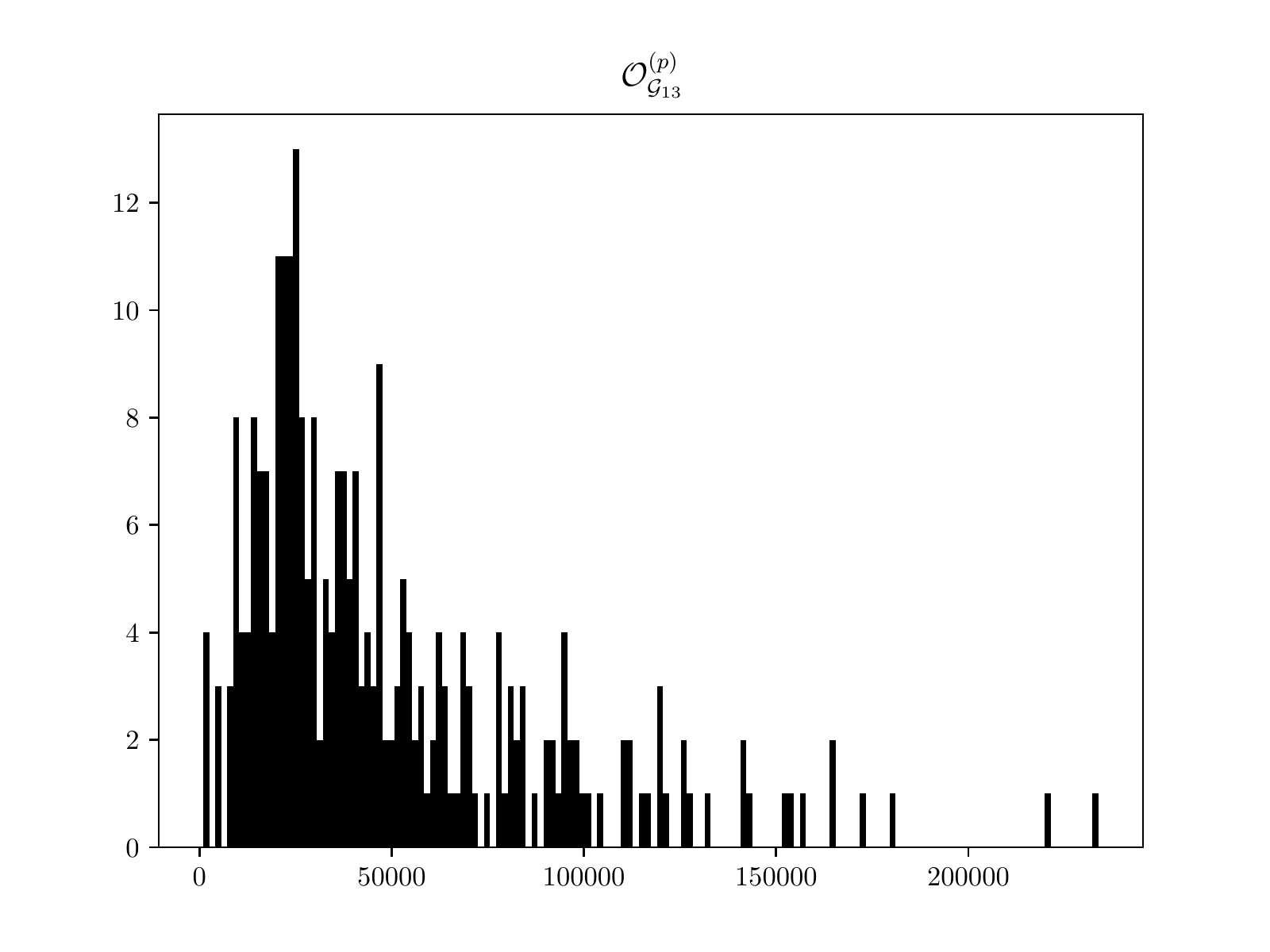}
\caption{Histogram for $\cO^{(p)}_{\cG_{13}} = \sum_{i,j,k} M_{ii}M_{jk}.$ Data collected from the adjective data set, at dimension D=2000.}
\label{fig:Histo(p)G13}
\end{figure}

\begin{figure}[ht!]
\centering
\includegraphics[width=14cm,height=11cm]{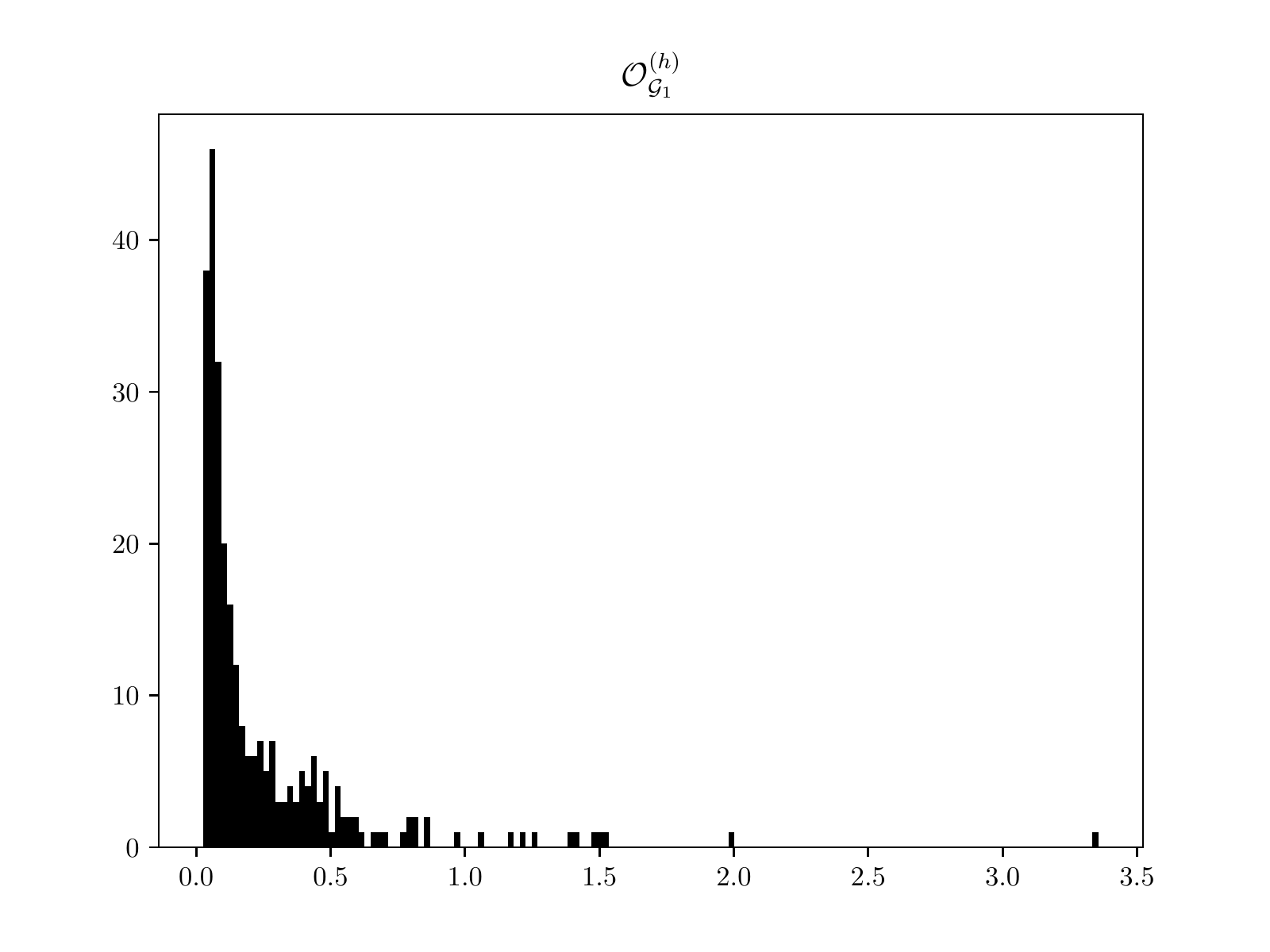}
\caption{Histogram for $\cO^{(h)}_{\cG_{1}} = $ $\sum_{ i } M_{  ii}^3 $. The majority of word expectation values are closely clustered around the mean. Data collected from the adjective data set, at dimension D=2000.}
\label{fig:Histo(h)G1}
\end{figure}

To describe precisely the typicality  properties of an observable, we can plot the histogram for the observable. For a given observable, we consider the range of its expectation values. We divide the range into a set of small bins, and we draw  a histogram where the heights of the rectangles in each bin are the numbers of 
words (i.e. adjectives or intransitive verbs) having their expectation values for the specified observable in that bin. These histograms can be constructed for both observables that parametrise the model, denoted with superscript $p$ (see figure \ref{fig:Histo(p)G13}) and also the higher order observables, denoted with superscript $h$. The histogram for a quadratic observable is 
given in Figure \ref{fig:Histo(p)G13} and a cubic observable in Figure \ref{fig:Histo(h)G1}. 
There is a significant diversity in the behaviour of the standard deviations of 
the ten observables $ \cO^{ (h)}_{ \cG_i }$ as a function of $D$. As observed at the beginning of this section, when these observables are divided by $ D^p$, we get standard deviations which go to zero 
as a function of $D$ in the large $ D $ region near $2000$. Interestingly, for the three of the observables $ \cO_{ \cG_1 }^{(h)}  , \cO^{(h)}_{ \cG_2 } , \cO^{(h)}_{ \cG_3 } $ the dispersions go to zero in this region even before dividing by $D^p$. Despite this diversity of behaviours in the standard deviations  as a function of $D$, the theoretical predictions based on the  Gaussian matrix models, work well for the whole range of observables considered, predicting the correct orders of magnitude for all of the observables.  
  


\subsection{ Theoretical predictions  for typicality  }

The permutation invariant Gaussian matrix model (PIGMM)  with fixed $ \mu , \Lambda $, determined 
by matching the linear and quadratic experimental averages, 
gives expectation values for  the higher order  polynomial invariants. 
By considering variations $ \delta \mu , \delta \Lambda $ to fit the 
experimental expectation values shifted by their standard deviations,
we can calculate the corresponding shifts in the theoretical expectation values. 
These shifts can be compared to the standard deviations in the higher order expectation values. 
This effectively involves using the PIGMM with a distribution of values of 
the $ \mu$ and  $\Lambda $ parameters, to predict both the expectation values and the dispersions of 
higher order observables.  It is interesting to note that physical models with random couplings are widely studied in condensed matter physics (e.g. \cite{Fischer}) and a class of these (SYK models \cite{SaYe,Kit}) have recently attracted interest  because of their potential links  \cite{WittenTensor}  to tensor model holography.

We now describe in more detail this prediction of  the dispersions of the higher order observables. 
We consider the Gaussian model parameterised by the $\mu_a$ parameters, for $ a \in \{ 1 , 2 \}$ and the eleven  parameters organised as matrix elements of $ \Lambda^{ V } $, for $ V \in \{ V_0 , V_H , V_2 , V_3 \}$.  We have an action $ S ( \mu_a , \Lambda^V )$ which determines the Gaussian measure. 
We have experiments parameterised by a binary choice - verbs or adjectives - and a choice of $D$. 
In  section \ref{sec:expect}, we have used, for each experiment, two linear 
 experimental expectation values and eleven quadratic expectation values. 
 We will refer to these thirteen 
 experimental expectation values as $ \langle \cO^{ (p)}_{ \cG } \rangle_{ EXPT} $. The superscript $p$ refers to the fact that these observables are used to parameterize the theoretical models. The subscript 
 $ \cG$ refers to the fact that the structure of the polynomial corresponds to a graph. This experimental input has been used to determine theoretical parameters $ \mu_a , \Lambda^V $.
 We have then used  these theoretical parameters to determine theoretical 
 cubic and quartic expectation values $\langle \cO^{(h)}_{ \cG } \rangle_{ TH }  $.  The superscript $h$ refers to observables of higher order than linear or quadratic. We tabulated the ratios
 \bea 
  \langle \cO^{(h)}_{ \cG } \rangle_{ TH }   \over \langle \cO^{(h)}_{ \cG } \rangle_{ EXPT} 
 \eea
for a number of these experiments. 

Using the histograms, for each parameterising observable we can determine 
a standard deviation  
\bea 
  ( \delta \cO^{(p)}_{ \cG })_{ EXPT} 
\eea
We can define positively shifted  variables
\bea
 \langle \cO^{(p)}_{ \cG } \rangle^{ +} =  \langle \cO^{(p)}_{ \cG } \rangle +  
 ( \delta \cO^{(p)}_{ \cG } )_{ EXPT} 
\eea 
we use  the linear system in section \ref{sec:linsys} to calculate shifted variables 
\bea 
\mu^{+ }_a ~ , ~ (\Lambda^{V} )^+   
\eea
These shifted variables are used to calculate theoretical shifted expectation values 
 $\langle \cO^{(h)}_{ \cG } \rangle^+_{ TH } $. We can repeat these steps with the negatively shifted 
 expectation values of $ \cO^{(p)  }_{ \cG }  $
 \bea 
  \langle \cO^{(p)}_{ \cG } \rangle^{ - } =  \langle \cO^{(p)}_{ \cG } \rangle -    ( \delta \cO^{(p)}_{ \cG })_{ EXPT} 
 \eea
Using the equations in Section \ref{sec:linsys}, these lead to 
\bea 
\mu^{-  }_a ~ , ~ (\Lambda^{V} )^-  
\eea

The positively and negatively shifted parameters define shifted theoretical values for the 
higher order observables  
\bea 
\langle \cO^{ (h) }_{ \cG } \rangle^{\pm}_{THEO}  
\eea
using the equations  (Appendix \ref{App:theorcubquart} ) derived from the matrix model. 
 
 Using the positively shifted theoretical parameters, we can define a magnitude of theoretical shift in the higher order observables  
  \bea 
  ( \delta^+ \cO^{ (h)}_{ \cG } )_{THEO}
  = |   \langle \cO^{(h)~  }_{ \cG}  \rangle_{THEO}^{+} - \langle \cO^{(h)~  }_{ \cG } \rangle_{ THEO}  | 
  \eea
  Similarly the negatively shifted theoretical parameters define a magnitude of 
  theoretical shift 
  \bea 
   ( \delta^-\cO^{ (h)}_{ \cG } )_{THEO}
  = |   \langle \cO^{(h)~  }_{ \cG}  \rangle_{THEO}^{-} - \langle \cO^{(h)~  }_{ \cG } \rangle_{ THEO}  |  
  \eea
  A measure of the theoretically predicted shift  in  expectation value is taken as the  average 
  \bea 
 ( \delta \cO^{(h)~  }_{ \cG_i  } )_{ THEO }   = 
{ 1 \over 2 } (       ( \delta^-  \cO^{ (h)}_{ \cG } )_{THEO} +   ( \delta^+ \cO^{ (h)}_{ \cG })_{THEO}  ) 
  \eea
In these theoretical predictions of the dispersion for the higher order polynomial invariants , we have taken as experimental input 26 parameters (13 expectation values and 13 standard deviations for linear and quadratic observables) from the data, which are being used alongside the equations of the  permutation invariant Gaussian matrix model. 
 
 In section \ref{sec:expect} we were 
 using the expectation values for the higher order observables $ \cO_{ \cG }^{ h } $. 
 A more refined look considers histograms for each observable.  The mean value extracted from the histogram is the expectation value used earlier. The standard deviation of each histogram determines 
 a $ ( \delta \cO^{(h)} )_{ EXPT } $.  In the tables below, for a number of the experiments, we tabulate 
 \bea 
{     ( \delta \cO^{ (h)}_{ \cG } )_{THEO}  \over  ( \delta  \cO^{(h)}_{\cG}  )_{ EXPT }}
 \eea
 
Using the above methodology, the standard deviation ratios between theory and experiment, for the 
$10$ cubic/quartic observables $ \cO_{\cG_i }^{ (h)} $  were calculated and are provided in the tables below. 

\vskip.2cm 
\begin{center}
{\bf Cubic and Quartic standard deviation ratios for 13 parameter model: } 
\end{center} 

\par
{\bf Adjectives at D = 2000:}

\vskip.2cm 

\begin{center}
\begin{tabular}{ | c | c | c | c | c | }
\hline
Graph & $\mathcal{O}^{(h)}_{\mathcal{G}_i}$ & $(\delta\mathcal{O}^{(h)}_{\mathcal{G}_i})_{THEO}$ & $( \delta \mathcal{O}^{(h)}_{\mathcal{G}_i} )_{EXPT}$ & Ratio \\[0.5mm]
\hline
1 & $\sum_{i} (M_{ii})^3 $ & $1.92 \times 10^{-1}$ & $3.52 \times 10^{-1}$ & 0.55\\ 
\hline
2 & $\sum_{i,j}  (M_{ij})^3 $ & $9.22 \times 10^{-1}$ & 2.70 & 0.34 \\ 
\hline
3 & $\sum_{i,j,k}  M_{ij}M_{jk}M_{ki} $ & 1.70 & 6.48 & 0.26\\
\hline
4 & $\sum_{i,j,k}  M_{ij}M_{jj}M_{jk} $ & 53.5 & 71.8 & 0.74\\
\hline
5 & $\sum_{i,j,k,l}  M_{ij}M_{kk}M_{ll} $ & $3.65 \times 10^{6}$ & $4.26 \times 10^{6}$ & 0.86\\ 
\hline
6 & $\sum_{i,j,k,l}  M_{ij}M_{jk}M_{ll} $ & $4.92 \times 10^{4}$ & $5.15 \times 10^{4}$ & 0.95\\ 
\hline
7 & $\sum_{i,j,k,l,m}  M_{ij}M_{kl}M_{mm} $ & $5.99 \times 10^{7}$ & $7.02 \times 10^{7}$ & 0.85 \\ 
\hline
8 & $\sum_{i,j,k,l,m,n}  M_{ij}M_{kl}M_{mn} $ & $1.46 \times 10^{9}$ & $1.69 \times 10^{9}$ & 0.86 \\ 
\hline
9 & $\sum_{i_{1}...i_{7}} M_{i_{1}i_{2}}M_{i_{3}i_{4}}M_{i_{5}i_{6}}M_{i_{7}i_{7}} $ & $ 8.72 \times 10^{10}$ & $ 1.28 \times 10^{11}$  & 0.67  \\
\hline
10 & $\sum_{i_{1}...i_{8}} M_{i_{1}i_{2}}M_{i_{3}i_{4}}M_{i_{5}i_{6}}M_{i_{7}i_{8}} $ & $2.31 \times 10^{12}$ & $3.28 \times 10^{12}$ & 0.70 \\ 
\hline
\end{tabular}
\end{center}

\vskip.2cm 

\noindent 
{\bf Adjectives at D = 700:}

\vskip.2cm 

\begin{center}
\begin{tabular}{ | c | c | c | c | c | }
\hline
Graph & $\mathcal{O}^{(h)}_{\mathcal{G}_i}$ & $(\delta\mathcal{O}^{(h)}_{\mathcal{G}_i})_{THEO}$ &$( \delta \mathcal{O}^{(h)}_{\mathcal{G}_i} )_{EXPT}$ & Ratio \\[0.5mm]
\hline
1 & $\sum_{i} (M_{ii})^3 $ & 2.96 & 4.93 & 0.60\\ 
\hline
2 & $\sum_{i,j}  (M_{ij})^3 $ & 7.97 & 13.5 & 0.59 \\ 
\hline
3 & $\sum_{i,j,k}  M_{ij}M_{jk}M_{ki} $ & 7.66 & 20.5 & 0.37\\
\hline
4 & $\sum_{i,j,k}  M_{ij}M_{jj}M_{jk} $ & $1.09 \times 10^{2}$ & $1.31 \times 10^{2}$ & 0.84\\
\hline
5 & $\sum_{i,j,k,l}  M_{ij}M_{kk}M_{ll} $ & $2.41 \times 10^{6}$ & $2.94 \times 10^{6}$ & 0.82\\ 
\hline
6 & $\sum_{i,j,k,l}  M_{ij}M_{jk}M_{ll} $ & $3.04 \times 10^{4}$ & $3.24 \times 10^{4}$ & 0.94\\ 
\hline
7 & $\sum_{i,j,k,l,m}  M_{ij}M_{kl}M_{mm} $ & $1.35 \times 10^{7}$ & $1.67 \times 10^{7}$ & 0.81 \\ 
\hline
8 & $\sum_{i,j,k,l,m,n}  M_{ij}M_{kl}M_{mn} $ & $1.11 \times 10^{8}$ & $1.35 \times 10^{8}$ & 0.82 \\ 
\hline
9 & $\sum_{i_{1}...i_{7}} M_{i_{1}i_{2}}M_{i_{3}i_{4}}M_{i_{5}i_{6}}M_{i_{7}i_{7}} $ & $8.27 \times 10^{9}$ & $1.38 \times 10^{10}$  & 0.60  \\
\hline
10 & $\sum_{i_{1}...i_{8}} M_{i_{1}i_{2}}M_{i_{3}i_{4}}M_{i_{5}i_{6}}M_{i_{7}i_{8}} $ & $7.48 \times 10^{10}$ & $1.19 \times 10^{11}$ & 0.63 \\ 
\hline
\end{tabular}
\end{center}

\vskip.2cm 

\noindent 
{\bf Adjectives at D = 1300:}

\vskip.2cm 

\begin{center}
\begin{tabular}{ | c | c | c | c | c | }
\hline
Graph & $\mathcal{O}^{(h)}_{\mathcal{G}_i}$ &  $(\delta\mathcal{O}^{(h)}_{\mathcal{G}_i})_{THEO}$
 &  $( \delta \mathcal{O}^{(h)}_{\mathcal{G}_i} )_{EXPT}$& Ratio \\[0.5mm]
\hline
1 & $\sum_{i} (M_{ii})^3 $ & $6.13 \times 10^{-1}$ & 1.05 & 0.58\\ 
\hline
2 & $\sum_{i,j}  (M_{ij})^3 $ & 2.234 & 5.25 & 0.43 \\ 
\hline
3 & $\sum_{i,j,k}  M_{ij}M_{jk}M_{ki} $ & 3.36 & 10.2 & 0.33\\
\hline
4 & $\sum_{i,j,k}  M_{ij}M_{jj}M_{jk} $ & 76.3 & 98.9 & 0.77\\
\hline
5 & $\sum_{i,j,k,l}  M_{ij}M_{kk}M_{ll} $ & $3.54 \times 10^{6}$ & $4.17 \times 10^{6}$ & 0.85\\ 
\hline
6 & $\sum_{i,j,k,l}  M_{ij}M_{jk}M_{ll} $ & $4.68 \times 10^{4}$ & $5.07 \times 10^{4}$ & 0.92\\ 
\hline
7 & $\sum_{i,j,k,l,m}  M_{ij}M_{kl}M_{mm} $ & $3.94 \times 10^{7}$ & $4.68 \times 10^{7}$ & 0.84 \\ 
\hline
8 & $\sum_{i,j,k,l,m,n}  M_{ij}M_{kl}M_{mn} $ & $6.2 \times 10^{8}$ & $7.33 \times 10^{8}$ & 0.85 \\ 
\hline
9 & $\sum_{i_{1}...i_{7}} M_{i_{1}i_{2}}M_{i_{3}i_{4}}M_{i_{5}i_{6}}M_{i_{7}i_{7}} $ & $4.34 \times 10^{10}$ & $6.59 \times 10^{10}$  & 0.66  \\
\hline
10 & $\sum_{i_{1}...i_{8}} M_{i_{1}i_{2}}M_{i_{3}i_{4}}M_{i_{5}i_{6}}M_{i_{7}i_{8}} $ & $7.35 \times 10^{11}$ & $1.09 \times 10^{12}$ & 0.67 \\ 
\hline
\end{tabular}
\end{center}

For half the observables, the prediction agrees with the data at a level above $70 \%$, the best ratios between theoretical and experimental standard deviations reaching $ 95 \%$, while the worst are at $26 \%$. Considering that the best ratios (for expectation values) obtained with the 5-parameter model 
were at $ 57 \%$ and the worst at $0.6 \%$, this can be considered another significant success of the thirteen parameter models. Another way to understand the range of ratios is to compare with a simple  Gaussian random walk model. Given that each of the observables involved a sum over a number of indices ranging from $1$ to $D$, the number of terms in each observable is $D^p$, where $p$ ranges from $1$ ( for 
$ \cO_{ \cG_1}  $ ) to $ 8$ for $ \cO_{ \cG_{10}} $. A simple random walk model for the dispersions in the first table is $ D^{p/2} \sigma $ for some constant $ \sigma $. Fixing $ \sigma $ to match exactly the last dispersion, we find
\bea 
\sigma = 0.205 
\eea
and  the following list of ratios for $ { \sigma D^{ p/2} / (\delta \cO^{ h}_{ \cG_i } })_{EXPT}  $ : 
\bea 
\{    26  , 151  , 2830 , 255 , 0.19 , 15.9 , 0.52 , 0.97 , 0.57 , 1\} 
\eea
The comparison  of the range  $ 0.19$ to $ 2830 $ of these ratios  to the range $ 26 \%  - 95 \% $ 
from the 13-parameter PIGMM, with random couplings, is another way to see the effectiveness of our theoretical framework for predicting the dispersions of the observables. 

\section{ Matrix model characteristics and correlations of word rankings  } 
\label{sec:rankings}

The inspection of the data on Gaussianity and typicality of the observables allows us to rank these observables in terms of how alike they are. For example $ \cO^{(h)}_{ \cG_2} , \cO^{(h)}_{ \cG_3} $ are very similar, while $ \cO^{(h)}_{ \cG_9} , \cO^{(h)}_{ \cG_{10}} $ are very similar to each other. 

In the course of developing the theory/experiment comparisons for the typicalities of the observables, we have made use of the histograms for these observables. These histograms are built by dividing the range of  values $  ( \cO^{(h)}_{ \cG } )_{EXPT}   $ into a number of bins and depicting in terms of vertical bars the multiplicity of words which have the evaluation of their 
 $  (\cO^{ (h)})_{EXPT}  $ in each bin. The dataset for adjectives has a total of 273 adjectives which fall in the various bins. 
 
 A refined look at each observable can be used to produce a ranked list of the adjectives, using the value of the observable as a ranking criterion, for example listing the adjective with the smallest 
 $ ( \cO^{(h)}_{ \cG })_{EXPT}  $ first and the one with the highest $ ( \cO^{ (h)}_{\cG} )_{EXPT} $ last.  
 
 Many computational tasks in distributional semantics work with ranked lists of words and compare their degrees of correlation. The main task here  is  ranking pairs of strings of words that are semantically related to each other. For example, \emph{SimLex-999} is a dataset that  quantifies the degree of similarity or relatedness  of 999 pairs of words,  such as  (cup, mug) and (cup, coffee). It   includes adjectives, nouns and verb pairs. Each pair is assigned a set of rankings as  judged by numerous human annotators and as predicted by different models. A degree of correlation is computed between rankings of different annotators and a set of different  models.  The models that better correlate with human annotations are returned as the ``better predicting" models. Often, the human annotators are also correlated with each other in order to find out how much do they agree with each other and to compute an \emph{inter-annotator agreement}.  These datasets have often  been specialised to only contain specific grammatical structures, e.g.  adjective noun phrases, as in \cite{MitLap2010},  which contains pairs of adjective noun phrases such as (last number, vast majority) together with gold-standard human similarity judgements. We also have the sentence similarity datasets mentioned in the previous section on Typicality. Adjective and verb similarity datasets, also mentioned in the section on Typicality,  are other examples. A further  slightly  different  task is inspired by  the dataset of \cite{Vec2015}, which  consists of  a set of unobserved acceptable phrases such as ``ethical statute"  and a set of  of deviant phrases such as ``cultural acne". The task is here to measure how well can different models distinguish between these two different pairs. A future direction of our project is to find out whether our model can predict and be applicable to any of these tasks.  
 

 As a first step in this direction, we investigate whether the patterns observed in the matrix model characteristics of the different observables are also reflected in the  properties of the ranked list for the observables. 
If two observables are very similar  in terms of  matrix model characteristics, do they produce very similar ranked lists ?

We have investigated this question by comparing the Spearman $\rho$ for the four observables using the adjectives data set at dimension $D=2000$. 
We find that $ \{ \cO_{ \cG_2} , \cO_{ \cG_3} , \cO_{ \cG_{9} } , \cO_{ \cG_{ 10} }  \}$ naturally split into two pairs which have excellent correlation : namely $ \{ \cO_{ \cG_2} , \cO_{ \cG_3} | \cO_{ \cG_9} , \cO_{ \cG_{ 10 } } \}$.  The correlation values are displayed in the following table:

\begin{table}[h]
\begin{center}
\begin{tabular}{| c | c | c |}
\hline
Observables Compared & Spearman $\rho$ & p-value\\
\hline
$\cO_{\cG_{9}}$ and $\cO_{\cG_{10}}$ & 0.97 & $1.20 \times 10^{-161}$\\
\hline
$\cO_{\cG_{2}}$ and $\cO_{\cG_{3}}$ & 0.88 & $4.57 \times 10^{-91}$\\
\hline
$\cO_{\cG_{2}}$ and $\cO_{\cG_{9}}$ & 0.81 & $8.72 \times 10^{-64}$\\
\hline
$\cO_{\cG_{3}}$ and $\cO_{\cG_{9}}$ & 0.80 & $2.91 \times 10^{-62}$\\
\hline
$\cO_{\cG_{2}}$ and $\cO_{\cG_{10}}$ & 0.66 & $2.11 \times 10^{-35}$\\
\hline
$\cO_{\cG_{3}}$ and $\cO_{\cG_{10}}$ & 0.66 & $3.69 \times 10^{-35}$ \\
\hline
\end{tabular}
\end{center}
\caption{Table shows the Spearman correlation coefficient and associated p-value for pairs of observable lists.}
\label{tab:spearman_rho}
\end{table}

Another way of comparing two ranked lists  $ \cL^{ (1)} , \cL^{ (2)} $ is to use two-dimensional correlation  plot. 
We can write the elements of the first list as $ \cL^{ (1)} = \{ L_1 , L_2 , \cdots L_n \}$. 
In the second list $ \cL^{ (2)} $,
 suppose $L_1 $ appears in position $ i_1 $, $L_2$ in position $i_2$ etc. 
We can plot on the $x-y$ plane, the points $ ( 1, i_1 ) , ( 2 , i_2 ) , \cdots , ( n , i_n)$. 

If the two ranked lists are identical, then these points fall on s straight line of gradient $1$. 
How close  the plots for two lists are to a straight line of gradient $1$ can be used as a visual estimate of their degree of similarity. The correlation plots for the significant pairs of lists produced from 
$  \{ \cO_{ \cG_2} , \cO_{ \cG_3} | \cO_{ \cG_9} , \cO_{ \cG_{ 10 } }  \}$ are shown below with the remaining plots given in appendix \ref{App:Spearmanplots}.

\begin{figure}[htbp!]
\centering
\includegraphics[width=11cm,height=9cm]{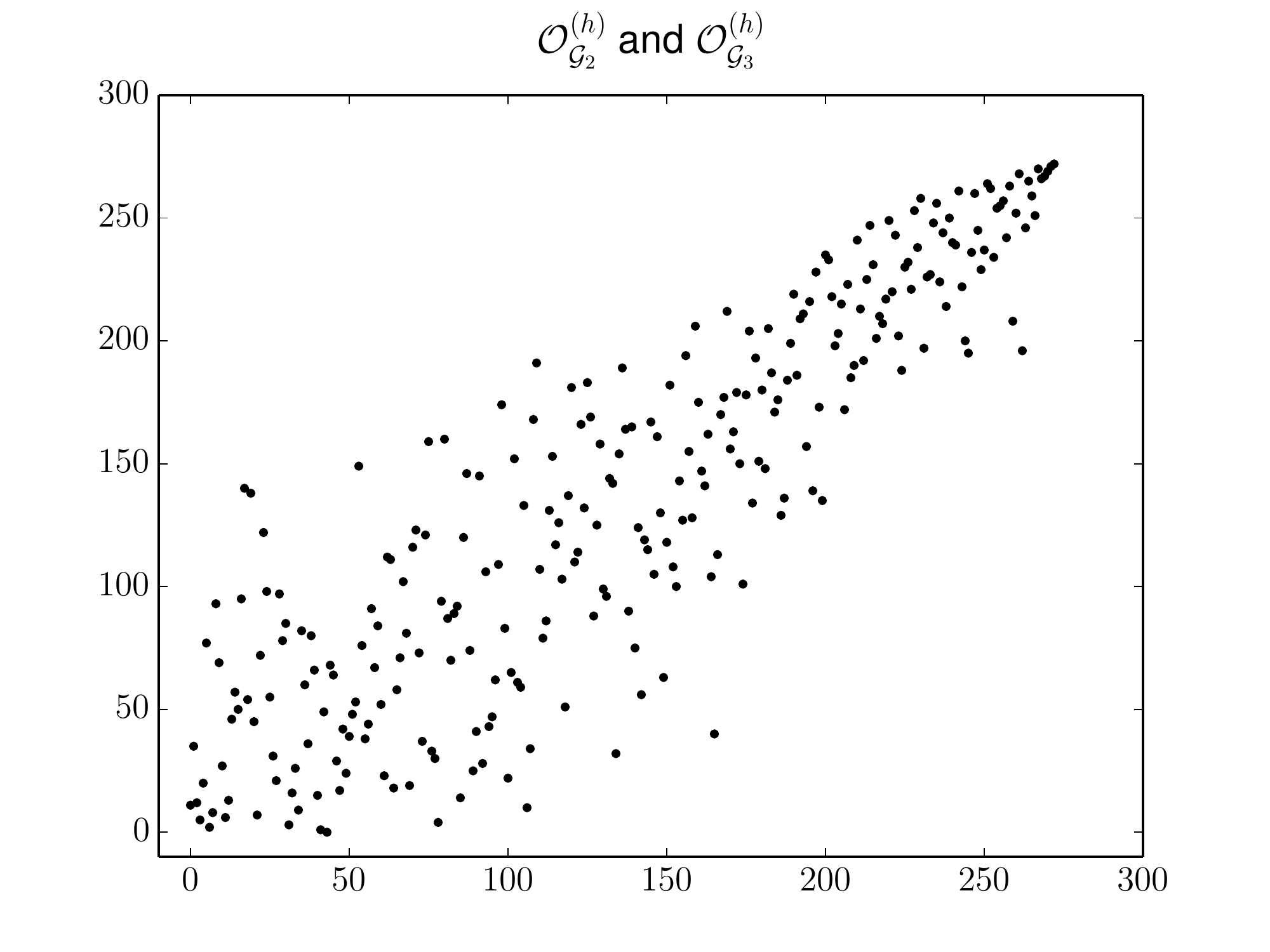}
\caption{Rank correlation plot corresponding to graph 2 and 3 observables}
\label{fig:O(h)G1vsO(h)G10}
\end{figure}

\begin{figure}[h!]
\centering
\includegraphics[width=11cm,height=9cm]{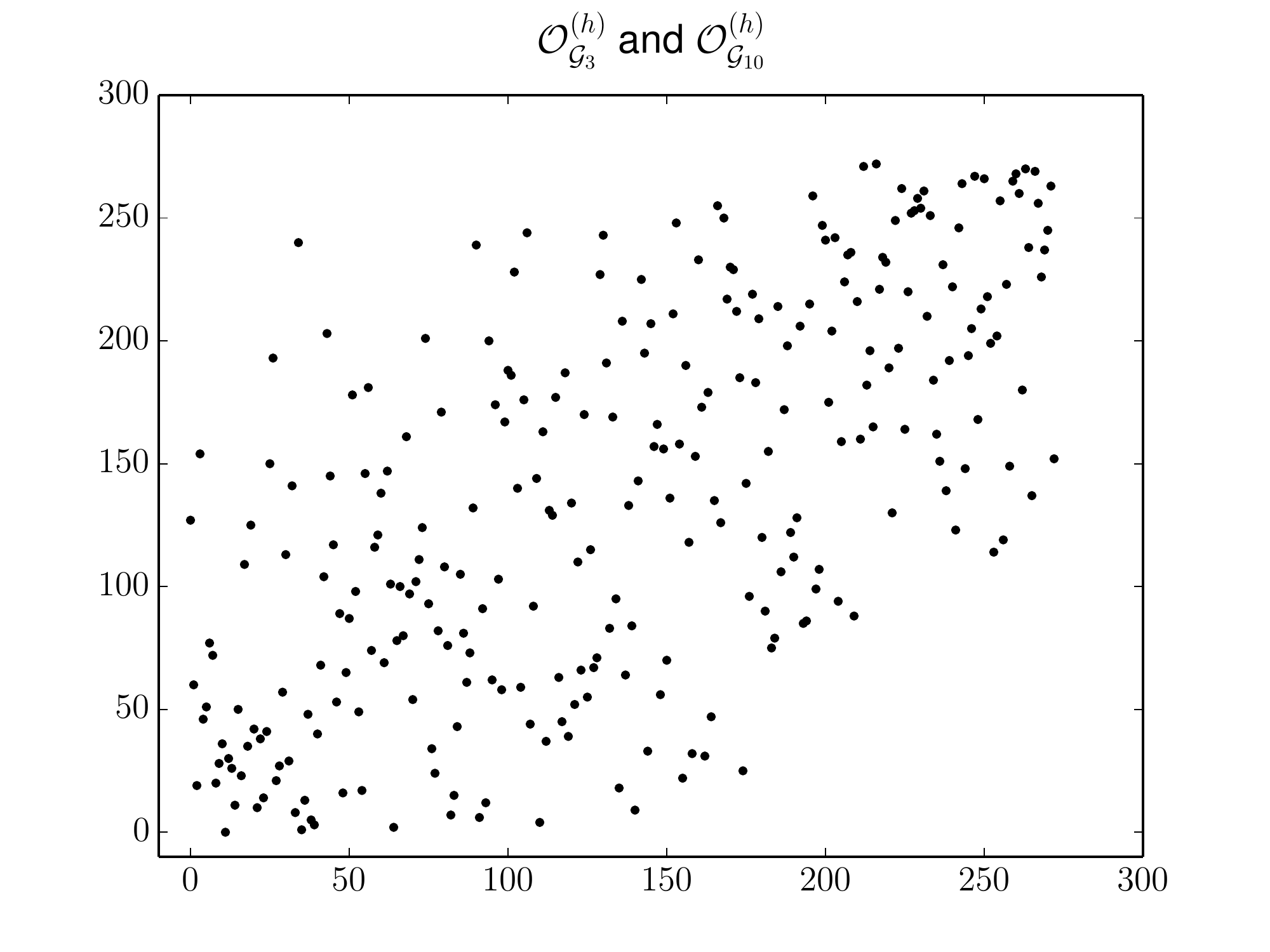}
\caption{Rank correlation plot corresponding to graph 3 and 10 observables}
\label{fig:O(h)G3vsO(h)G10}
\end{figure}

\begin{figure}[h!]
\centering
\includegraphics[width=11cm,height=9cm]{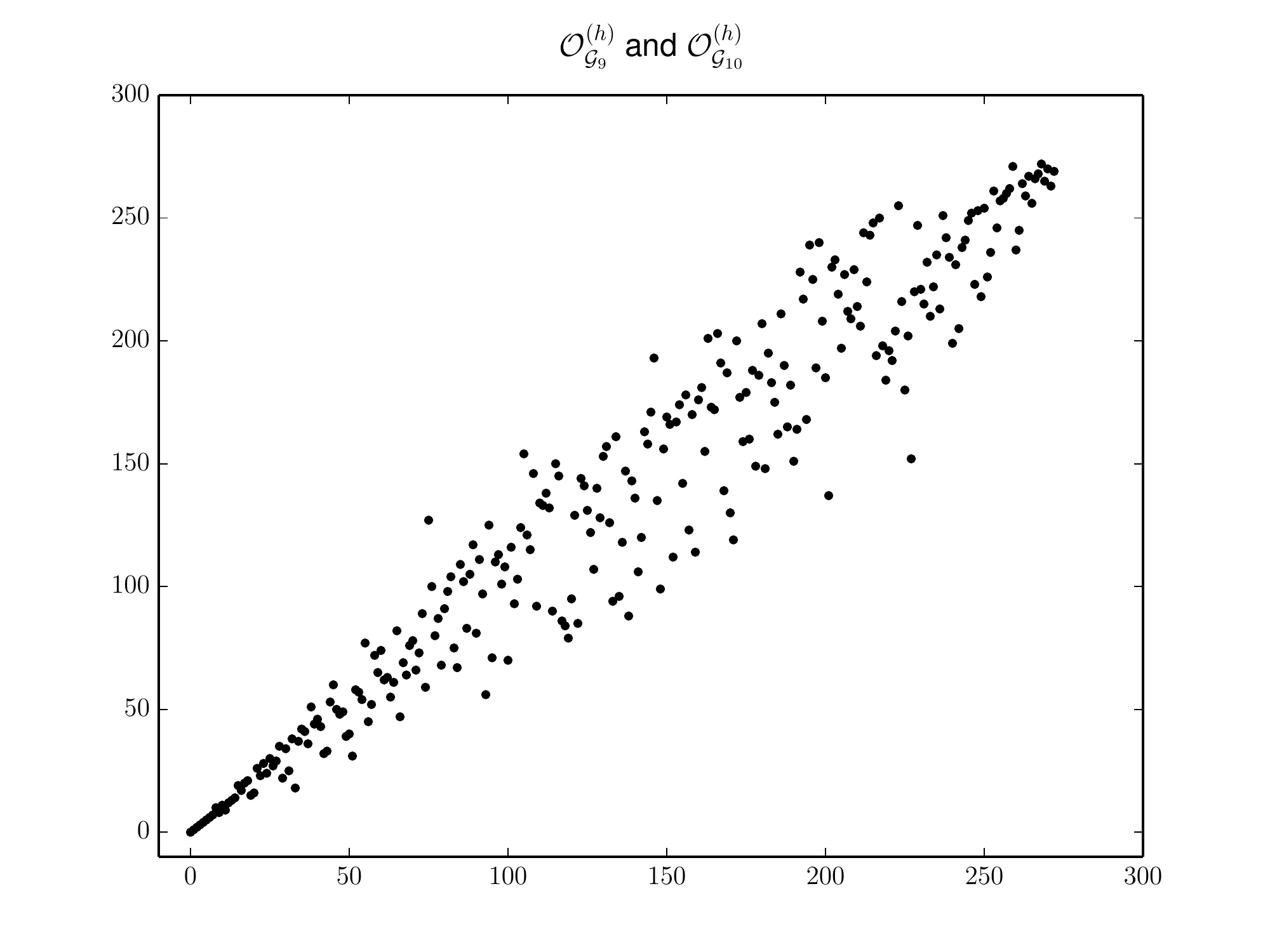}
\caption{Rank correlation plot corresponding to graph 9 and 10 observables}
\label{fig:O(h)G9vsO(h)G10}
\end{figure}

It is evident again, in agreement with the Spearman analysis, that the most well correlated pairs are 
$ \{ \cO_{ \cG_2} , \cO_{ \cG_3} \}  $ and  $ \{  \cO_{ \cG_9} , \cO_{ \cG_{ 10 } }  \}$

We conjecture that a systematic study of the degree of similarity between the matrix model characteristics of observables will show that these are indeed very well correlated with the lists. 

These regularities in the matrix model characteristics and ranked lists associated with observables are observational properties of the data. Is there a theoretical prediction of this property ? Given the usefulness of ranked lists in the tasks of distributional semantics, are these regularities an avenue towards applications of the matrix model perspective as a tool in facilitating concrete tasks in computational linguistics ? 

\section{ Summary and Outlook }

Matrix models have had widespread success in capturing universal characteristics of randomness
in diverse types of complex systems \cite{Wigner,Dyson,GMW98,Been1997,EY2013,RMTMN0503}. 
The program of Linguistic Matrix theory (LMT)  \cite{LMT,PIGMM}  follows the same philosophy and aims to characterise universal features of the randomness in the matrices/tensors constructed in type-driven compositional distributional semantics. Concretely it postulates Gaussianity in the expectation values of permutation 
invariant polynomial functions of matrices associated with adjectives, or intransitive verbs. 
In this paper, we have found high levels of success in the predictions of Gaussianity for a significant number of observables. These tests of Gaussianity have been formulated both for the
expectation values of observables, as well as the standard deviations of the observables. 
 Another strong piece of evidence in favour of the Gaussianity hypothesis is 
that, in all the experiments we have done, the theoretical parameters extracted are compatible with convergent Gaussian measures. These high levels of success  show that the Gaussianity  hypothesis is fundamentally sound, and this raises a number of questions for further investigation. 

\begin{itemize} 

\item  For a small number of observables, $  \cO_{ \cG_2 } , \cO_{ \cG_3 } $ in Table \ref{tab:spearman_rho}, 
the theory/experiment ratios are noticeably smaller than the others. One possibility is that certain 
justifiable improvements in the algorithms for constructing the matrices can increase these ratios. 
For example, in the present construction of the noun vectors ( which are fed into a linear regression method to produce adjective matrices), it can happen that the list of nouns has some overlap with the 
list of context words. When this happens, there is an issue of how to count the frequency of 
proximity of a word with itself. The present algorithm uses a reasonable, but perhaps non-unique, choice for handling these cases. A modification of the algorithm would exclude these cases from the construction of word vectors, and investigate the resulting matrices.  

\item If it turns out that $ \cO_{ \cG_2 } , \cO_{ \cG_3 } $ really are less Gaussian than the remaining observables ( as in the table \ref{tab:spearman_rho}), after any reasonable changes in the construction algorithms for the matrices,  then  we may ask how to modify the matrix model so as to increase the accuracy of prediction for these expectation values. Perturbations of the Gaussian model by  adding these specific observables as perturbations would be a natural guess.

\item Can we get comparable or higher levels of success in predicting expectation values when using different matrix constructions? How universal are the statistical characteristics we are finding?  Different constructions have been used in the computational side to produce the linguistic matrices, e.g.  algorithms such as  linear regression \cite{BZ2010}, multi-step linear regression \cite{GDZSB13},  and neural networks, e.g. the extensions of the  hierarchal softmax  algorithm of the Word2Vec  model  of \cite{MSCCD2013}, developed in \cite{MC2015}. The main ideas behind these constructions is the same: they all explore the original intuitions of Firth and Harris, that we can  use the context of the words and the degrees of similarities between them to build matrices for words with functional types. As the methods behind these constructions advance, the matrices become denser and learn to perform better in tasks that they are trained on.

\item In section  \ref{sec:rankings} we have investigated rankings of words directly associated with 
the observables.  The idea of considering rankings was motivated by uses of word and phrase rankings in AI tasks, as discussed in detail in section \ref{sec:typicality}. Relating the rankings of word and phrase similarity tasks to the rankings associated with the observables of the matrix model (section \ref{sec:rankings}) is a very interesting avenue for further investigation.

\item It will be interesting to compute the matrix model characteristics - the $\mu , \Lambda$ parameters, the theory/experiment ratios for higher order observables, and their standard deviations - for other linguistic corpora, e.g. specialising to particular genres of literature, different  domains (e.g. news articles)  and modes (e.g audio and video)   of content, or using other languages than English. This will be a way to identify which of  the matrix model characteristics are universal and which are corpus-dependent. 

\end{itemize}

The conventional applications of distributional semantics  in AI 
focus on structural  aspects of the data related to the meanings assigned by humans to words. 
LMT focuses on the characteristics of the randomness, successfully predicts some of these characteristics to high accuracy, and demonstrate simple patterns in the success rates in terms of structures
of the observables as encoded in graphs. A very interesting conceptual question is: How do structure and randomness interface in distributional semantics ? The observables and experiments in this paper provide some tools for investigating this question. Mathematical perspectives on the broad question of interfaces between structure and randomness are discussed in \cite{tao-structure-randomness}. 

 Natural language is a very interesting natural and complex system, amenable through universal perspectives based in matrix theories, to ideas from theoretical  physics. We have so far used ideas from theoretical physics to identify regularities in the randomness 
present in language. An interesting question for the future is whether the characterization 
of the universality classes of randomness existing in language holds some lessons for theoretical physics, in its quest to understand the complex natural system that is the universe.  

\vskip2cm 

 \begin{centerline} 
{\bf Acknowledgments} 
\end{centerline} 
\vskip.4cm 

The research of SR is supported by the  STFC Grant ST/P000754/1, ``String Theory, Gauge Theory, and Duality''; and by a Visiting Professorship at the University of  the Witwatersrand, funded by a Simons Foundation grant (509116) awarded to the Mandelstam Institute for Theoretical Physics at Wits. 
MS is supported by Royal Academy of Engineering Industrial Scheme Fellowship, IF\_192058. 
 We thank for useful discussions David Arrowsmith, Andreas Brandhuber, Robert de Mello Koch, Chris Hull, Hally Ingram,  Dimitri Kartsaklis,  Costis Papageorgakis, Gabriele Travaglini, Chris White.  

\appendix

\section{Theoretical equations for cubic and quartic expectation values}
\label{App:theorcubquart}

\begin{align*}
    \sum_{i} \langle M_{ii}^{3} \rangle &= 3\bigg( \frac{1}{D} \widetilde{\mu}_{1} + \frac{\sqrt{(D-1)}}{D} \widetilde{\mu}_{2} \bigg) \times \Bigg( \frac{1}{D}(\Lambda_{V_{0}}^{-1})_{11} + \frac{(D-1)}{D}(\Lambda_{V_{0}}^{-1})_{22} + 2 \frac{\sqrt{(D-1)}}{D} (\Lambda_{V_{0}}^{-1})_{12} \\
    & + \frac{(D-1)}{D}(\Lambda_{H}^{-1})_{11} + \frac{(D-1)}{D}(\Lambda_{H}^{-1})_{22} + \frac{(D-1)(D-2)}{D} (\Lambda_{H}^{-1})_{33} + 2\frac{(D-1)}{D} (\Lambda_{H}^{-1})_{12} \\
    & + 2\frac{(D-1)}{D} \sqrt{(D-2)} (\Lambda_{H}^{-1})_{13} + 2\frac{(D-1)}{D} \sqrt{(D-2)} (\Lambda_{H}^{-1})_{23} + \frac{1}{3D} \big(\widetilde{\mu}_{1} + \sqrt{(D-1)}\widetilde{\mu}_{2} \big)^{2} \Bigg).
\end{align*}

\begin{align*}
    \sum_{i,j} \langle M_{ij}^{3} \rangle &= \frac{\widetilde{\mu}_{1}^{3}}{D} + \frac{3}{D}\widetilde{\mu}_{1}\widetilde{\mu}_{2}^{2} + \frac{(D-2)}{D \sqrt{(D-1)}} \widetilde{\mu}_{2}^{3} + \frac{3 \widetilde{\mu}_{1}}{D} (\Lambda_{V_0}^{-1})_{11} + \frac{3 \widetilde{\mu}_{1}}{D} (\Lambda_{V_0}^{-1})_{22} + \frac{3 \widetilde{\mu}_{1}}{D} (D-1) (\Lambda_{H}^{-1})_{22} \\
    & + \frac{3 \widetilde{\mu}_{1}}{D} (D-1) (\Lambda_{H}^{-1})_{33} + \frac{3 \widetilde{\mu}_{1}}{D} (D-1) (\Lambda_{H}^{-1})_{11} + \frac{3 \widetilde{\mu}_{1}}{D} \frac{D(D-3)}{2} (\Lambda_{V_2}^{-1}) \\
    & + \frac{3 \widetilde{\mu}_{1}}{D} \frac{(D-1)(D-2)}{2} (\Lambda_{V_3}^{-1}) + \frac{3 \widetilde{\mu}_{2}(D-2)}{D \sqrt{(D-1)}} (\Lambda_{V_0}^{-1})_{22} + \frac{6 \widetilde{\mu}_{2}}{D} (\Lambda_{V_0}^{-1})_{12} \\
    & + \frac{3 \widetilde{\mu}_{2}(D-3)\sqrt{(D-1)}}{D} (\Lambda_{H}^{-1})_{33} + \frac{6 \widetilde{\mu}_{2} \sqrt{(D-1)}}{D} (\Lambda_{H}^{-1})_{12} + \frac{6 \widetilde{\mu}_{2} \sqrt{(D-1)(D-2)}}{D}(\Lambda_{H}^{-1})_{13} \\
    & + \frac{6 \widetilde{\mu}_{2} \sqrt{(D-1)(D-2)}}{D}(\Lambda_{H}^{-1})_{23} + \frac{3 \widetilde{\mu}_{2} (3D - D^{2})}{2D \sqrt{(D-1)}} (\Lambda_{V_2}^{-1}) - \frac{3 \widetilde{\mu}_{2} (D-2) \sqrt{(D-1)}}{D} (\Lambda_{V_3}^{-1}).
\end{align*}

\begin{align*}
    \sum_{i,j,k} \langle M_{ij}M_{jk}M_{ki} \rangle &=  \widetilde{\mu}_{1}^{3} + \frac{\widetilde{\mu}_{2}^{3}}{\sqrt{(D-1)}} + 3 \widetilde{\mu}_{1}(\Lambda_{V_0}^{-1})_{11} + 3 \widetilde{\mu}_{1}(D-1)(\Lambda_{H}^{-1})_{12} + \frac{3\widetilde{\mu}_{2}}{\sqrt{(D-1)}} (\Lambda_{V_{0}}^{-1})_{22} \\
       & + 3\widetilde{\mu}_{2} \sqrt{(D-1)}(\Lambda_{H}^{-1})_{33} + 3\widetilde{\mu}_{2} \sqrt{(D-1)}(\Lambda_{H}^{-1})_{12} + 3 \widetilde{\mu}_{2} (\Lambda_{2}^{-1}) \frac{D(D-3)}{2\sqrt{(D-1)}} \\
       & - 3 \widetilde{\mu}_{2} (\Lambda_{3}^{-1}) \frac{(D-2)\sqrt{(D-1)}}{2}.
\end{align*}

\begin{align*}
    \sum_{i,j,k} \langle M_{ij} M_{jj} M_{jk} \rangle &= 3\widetilde{\mu}_{1}(\Lambda_{V_0}^{-1})_{11} + 2\widetilde{\mu}_{1}\sqrt{(D-1)}(\Lambda_{V_0}^{-1})_{12} + \widetilde{\mu}_{1}(D-1)(\Lambda_{V_H}^{-1})_{11} + \widetilde{\mu}_{1}(D-1)(\Lambda_{V_H}^{-1})_{22} \\
    & + 3\widetilde{\mu}_{1}(D-1)(\Lambda_{V_H}^{-1})_{12} + 2\widetilde{\mu}_{1}(D-1)\sqrt{(D-2)}(\Lambda_{V_H}^{-1})_{13} + \widetilde{\mu}_{2}\frac{(D-1)}{\sqrt{(D-2)}}(\Lambda_{V_0}^{-1})_{11} \\
    & + \widetilde{\mu}_{2}\frac{(D-1)^2}{\sqrt{(D-2)}}(\Lambda_{V_H}^{-1})_{12} + \widetilde{\mu}_{1}^3 + \widetilde{\mu}_{1}^{2}\widetilde{\mu}_{2}\sqrt{(D-1)}.
\end{align*}

\begin{align*}
     \sum_{i,j,k,l} \langle M_{ij} M_{kk} M_{ll} \rangle &= 3D\widetilde{\mu}_{1}(\Lambda_{V_0}^{-1})_{11} + 2D \sqrt{(D-1)} \widetilde{\mu}_{2}(\Lambda_{V_0}^{-1})_{11} + 2D\sqrt{(D-1)} \widetilde{\mu}_{1}(\Lambda_{V_0}^{-1})_{12} \\ 
     & + 2D(D-1)\widetilde{\mu}_{2}(\Lambda_{V_0}^{-1})_{12}  + D (D-1)\widetilde{\mu}_{1}(\Lambda_{V_0}^{-1})_{22} 
     + 2 D \sqrt{(D-1)}\widetilde{\mu}_{1}(\Lambda_{V_0}^{-1})_{12} \\
     & + D \widetilde{\mu}_{1} (\widetilde{\mu}_{1} + \sqrt{(D-1)}\widetilde{\mu}_{2})^{2}.
\end{align*}

\begin{align*}
    \sum_{i,j,k,l} \langle M_{ij}M_{jk}M_{ll} \rangle &= 3D\widetilde{\mu}_{1}(\Lambda_{V_0}^{-1})_{11} + D\sqrt{(D-1)}\widetilde{\mu}_{2}(\Lambda_{V_0}^{-1})_{11} + D(D-1)\widetilde{\mu}_{1}(\Lambda_{V_H}^{-1})_{12} \\ 
    & + D(D-1)\sqrt{(D-1)}\widetilde{\mu}_{2}(\Lambda_{V_H}^{-1})_{12} + 2D\sqrt{(D-1)}\widetilde{\mu}_{1}(\Lambda_{V_0}^{-1})_{12}                                          \\
    & + D\widetilde{\mu}_{1}^{3} + D\sqrt{(D-1)}\widetilde{\mu}_{1}^{2}\widetilde{\mu}_{2}.
\end{align*}

\begin{align*}
    \sum_{i,j,k,l,m} \langle M_{ij}M_{kl}M_{mm} \rangle
        & = D^{2} \widetilde{\mu}_{1}^{3} + D^{2}\widetilde{\mu}_{1}(\Lambda_{V_0}^{-1})_{11} + D^{2}\sqrt{(D-1)}\widetilde{\mu}_{1}^{2}\widetilde{\mu}_{2} + D^{2} \sqrt{(D-1)}\widetilde{\mu}_{2}(\Lambda_{V_0}^{-1})_{11} \\
        & + 2D^{2}\widetilde{\mu}_{1} (\Lambda_{V_0}^{-1})_{11} + 2D^{2}\sqrt{(D-1)}\widetilde{\mu}_{1}(\Lambda_{V_0}^{-1})_{12}.
\end{align*}

\begin{align*}
    \sum_{i,j,k,l,m,n} \langle M_{ij}M_{kl}M_{mn} \rangle = 3 \widetilde{\mu}_{1} D^{3}(\Lambda_{V_0}^{-1})_{11} + D^{3}\widetilde{\mu}_1^{3}.
\end{align*}

\begin{align*}
     \sum_{i_{1},i_{2},i_{3},i_{4},i_{5},i_{6},i_{7}}& \langle M_{i_{1}i_{2}}M_{i_{3}i_{4}}M_{i_{5}i_{6}}M_{i_{7}i_{7}} \rangle = 3D^{3} (\Lambda_{V_0}^{-1})_{11}^{2} + 3D^{3} \sqrt{(D-1)}(\Lambda_{V_0}^{-1})_{11}(\Lambda_{V_0}^{-1})_{12} \\
     & + 6D^{3} \widetilde{\mu}_{1}^{2} (\Lambda_{V_0}^{-1})_{11} + 3D^{3} \sqrt{(D-1)} \widetilde{\mu}_{1} \widetilde{\mu}_{2} (\Lambda_{V_0}^{-1})_{11} + D^{3}\sqrt{(D-1)}\widetilde{\mu}_{1}^{2} (\Lambda_{V_0}^{-1})_{12} +
     \\
     & + D^{3}\widetilde{\mu}_{1}^{4} + D^{3}\sqrt{(D-1)}\widetilde{\mu}_{1}^{3}\widetilde{\mu}_{2}.
\end{align*}

\begin{align*}
    \sum_{i_{1},i_{2},i_{3},i_{4},i_{5},i_{6},i_{7},i_{8}} \langle M_{i_{1}i_{2}}M_{i_{3}i_{4}}M_{i_{5}i_{6}}M_{i_{7}i_{8}} \rangle = 3D^{4} (\Lambda_{V_0}^{-1})_{11}^2 + 6D^{4}\widetilde{\mu}_{1}^{2} (\Lambda_{V_0}^{-1})_{11} + D^{4}\widetilde{\mu}_{1}^{4}.
\end{align*}

\section{Additional Spearman $\rho$ plots}
\label{App:Spearmanplots}
The remaining rank correlation plots associated to the Spearman $\rho$ calculations made in table \ref{tab:spearman_rho} are provided below. Collectively, all the plots in Section \ref{sec:rankings}
and below confirm the conclusion that $ \cO^{(h)}_{ \cG_{ 9} } , \cO^{(h)}_{ \cG_{ 10 } }$ have the best pairwise correlation of the associated ranked list of words, followed by the pair  
 $ \cO^{(h)}_{ \cG_{ 2} } , \cO^{(h)}_{ \cG_{ 3 } }$. This reflects the similarity in the matrix model characteristics between these pairs (theory/expt ratios for expectation values and standard deviations) 
 which are  visible in Sections 3,4,5.  

\begin{figure}[h!]
\centering
\includegraphics[width=11cm,height=9cm]{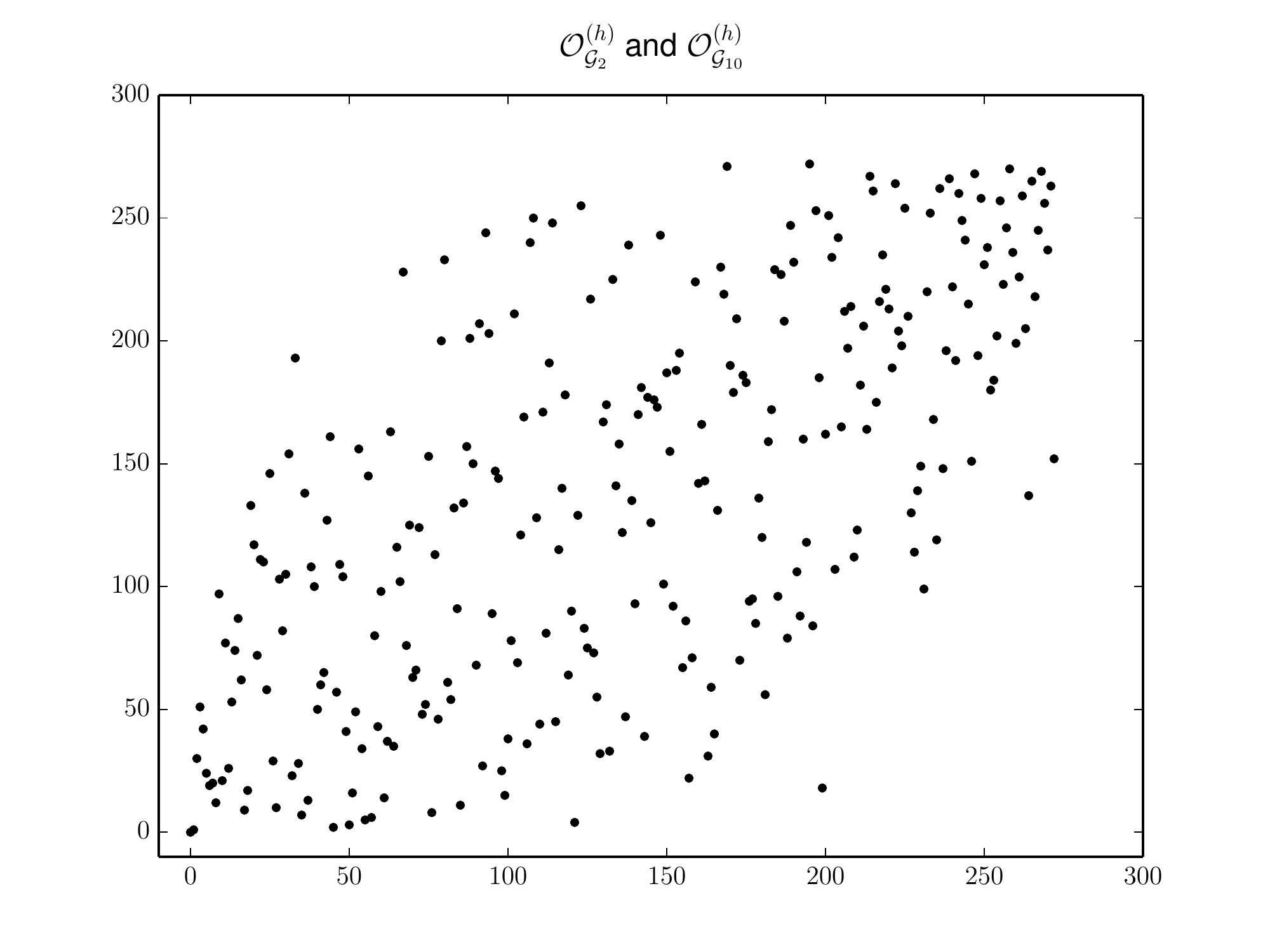}
\caption{Rank correlation plot corresponding to graph 2 and 10 observables}
\label{fig:O(h)G1vsO(h)G9}
\end{figure}

\begin{figure}[h!]
\centering
\includegraphics[width=11cm,height=9cm]{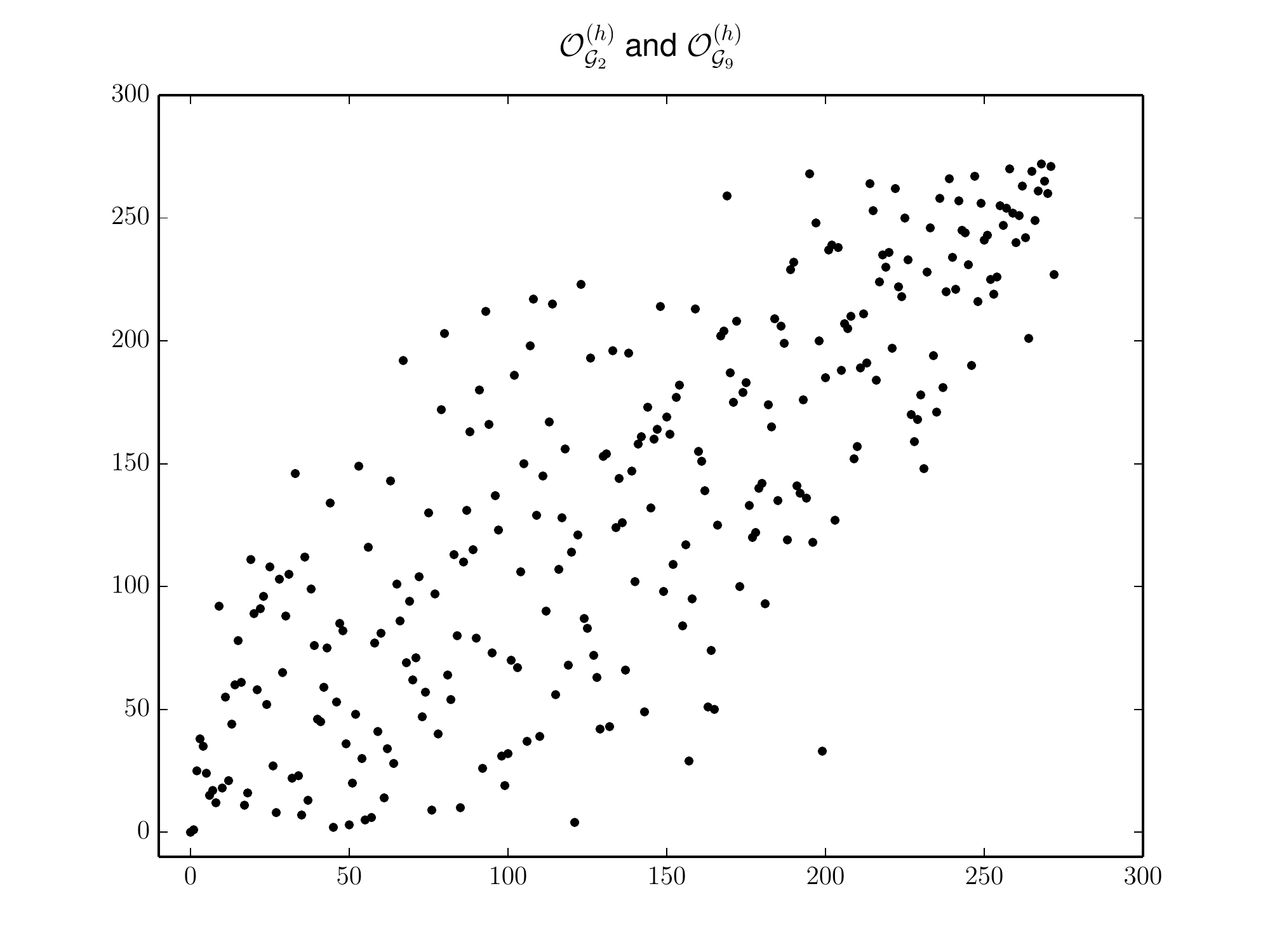}
\caption{Rank correlation plot corresponding to graph 2 and 9 observables}
\label{fig:O(h)G2vsO(h)G9}
\end{figure}

\begin{figure}[h!]
\centering
\includegraphics[width=11cm,height=9cm]{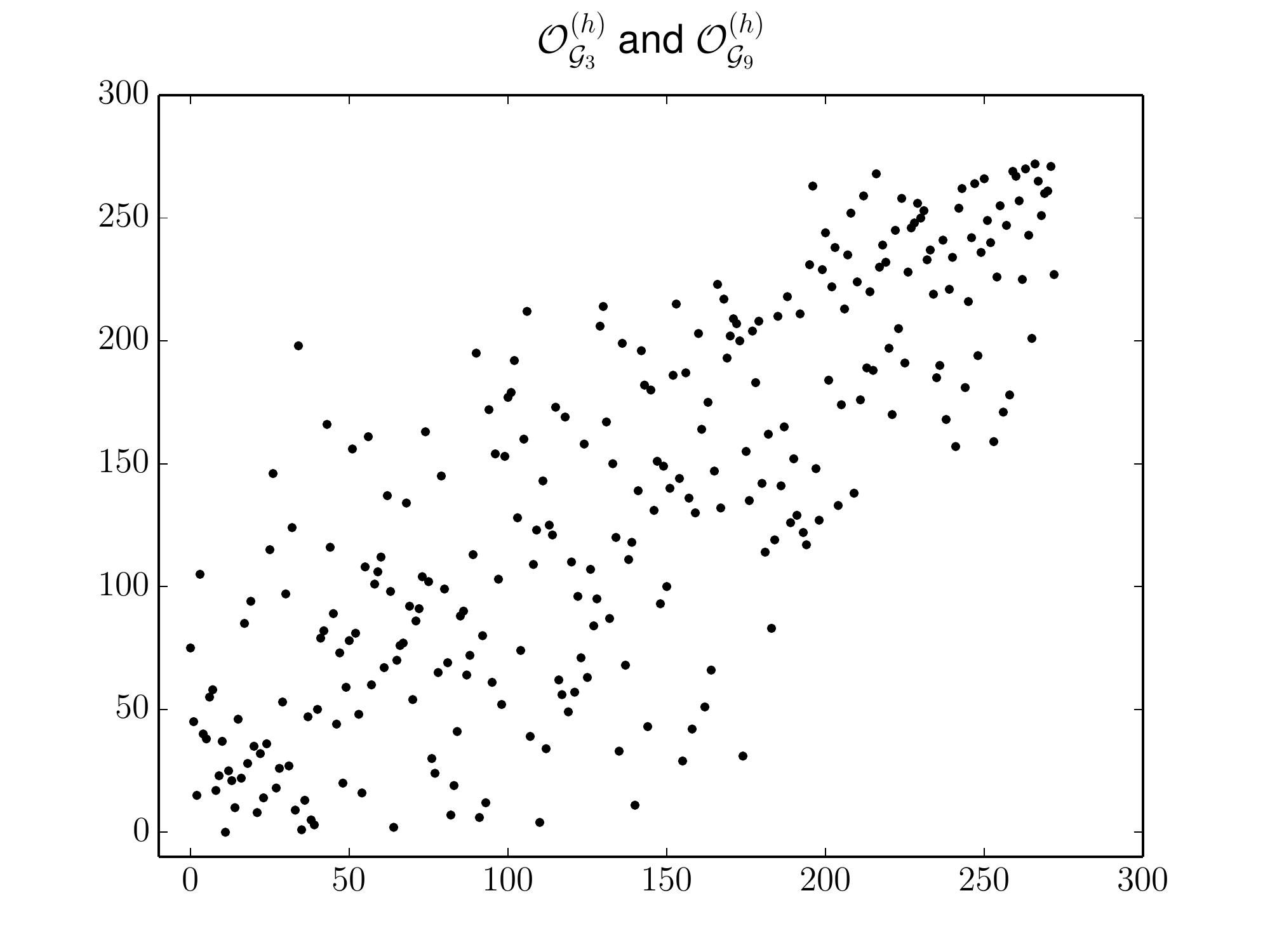}
\caption{Rank correlation plot corresponding to graph 3 and 9 observables}
\label{fig:O(h)G2vsO(h)G10}
\end{figure}

\newpage
\section{Graph diagrams for higher order observables}
\label{App:Graphdiagram}

\begin{figure}[!htbp]
  \captionsetup[subfigure]{labelformat=empty}
  \centering
	\begin{subfigure}[b]{.275\textwidth}
  		\centering
  		\includegraphics[width=.5\linewidth]{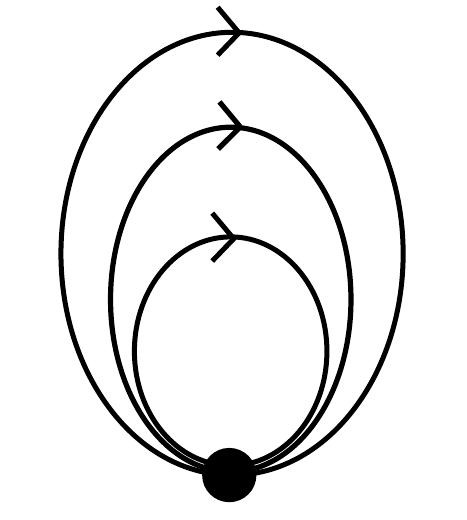}
 		 \caption{$\sum_{i} M_{ii}^{3}$}
  	\label{fig:Graph1}
	\end{subfigure}\hspace{-7mm}%
	\begin{subfigure}[b]{.25\textwidth}
 		 \centering
  		\includegraphics[width=.5\linewidth]{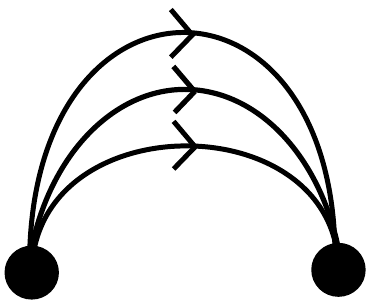}
  		\caption{$\sum_{i,j} M_{ij}^{3}$}
	 \label{fig:Graph2}
	\end{subfigure}%
	\begin{subfigure}[b]{.25\textwidth}
	  	\centering
 		 \includegraphics[width=.75\linewidth]{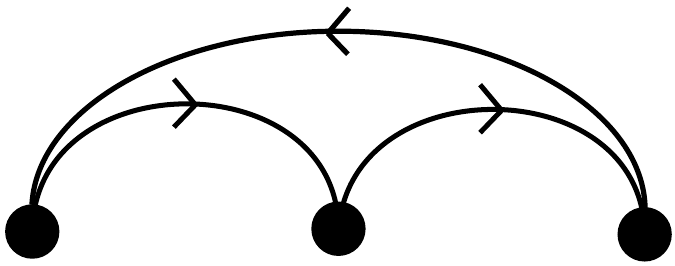}
	 	 \caption{$\sum_{i,j,k} M_{ij}M_{jk}M_{ki}$}
	 \label{fig:Graph3}
	\end{subfigure}%
	\begin{subfigure}[b]{.25\textwidth}
	  	\centering
 		 \includegraphics[width=.75\linewidth]{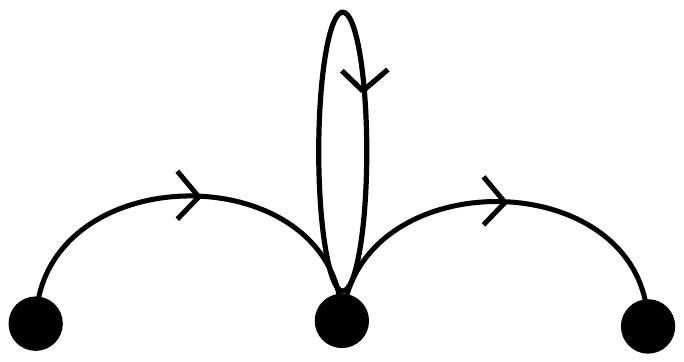}
	 	 \caption{$\sum_{i,j,k} M_{ij} M_{jj} M_{jk}$ }
	 \label{fig:Graph4}
	\end{subfigure}%
	\vspace{5mm}
	\begin{subfigure}[b]{.3\textwidth}
  		\centering
  		\includegraphics[width=0.8\linewidth]{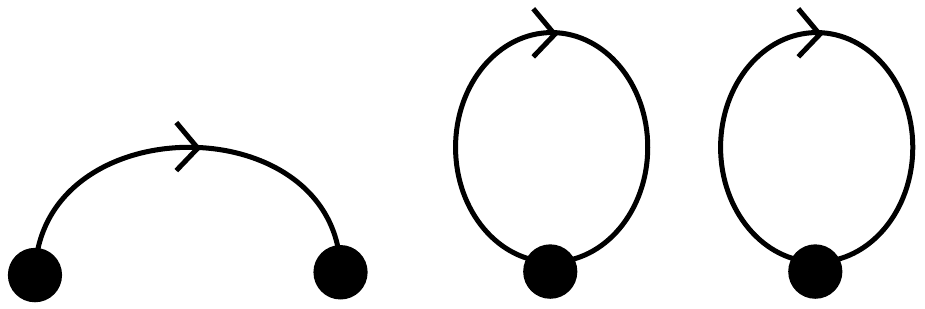}
 		 \caption{$\sum_{i,j,k,l} M_{ij} M_{kk} M_{ll}$}
  	\label{fig:Graph5}
	\end{subfigure}%
	\begin{subfigure}[b]{.3\textwidth}
 		 \centering
  		\includegraphics[width=0.8\linewidth]{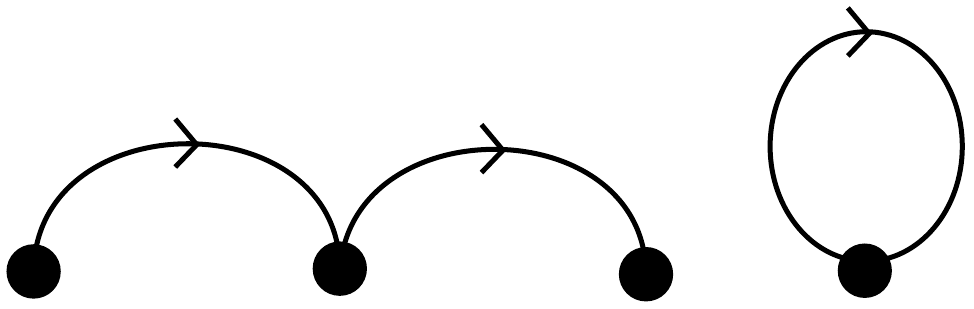}
  		\caption{$\sum_{i,j,k,l} M_{ij}M_{jk}M_{ll}$ }
	 \label{fig:Graph6}
	\end{subfigure}%
	\begin{subfigure}[b]{.3\textwidth}
	  	\centering
 		 \includegraphics[width=0.8\linewidth]{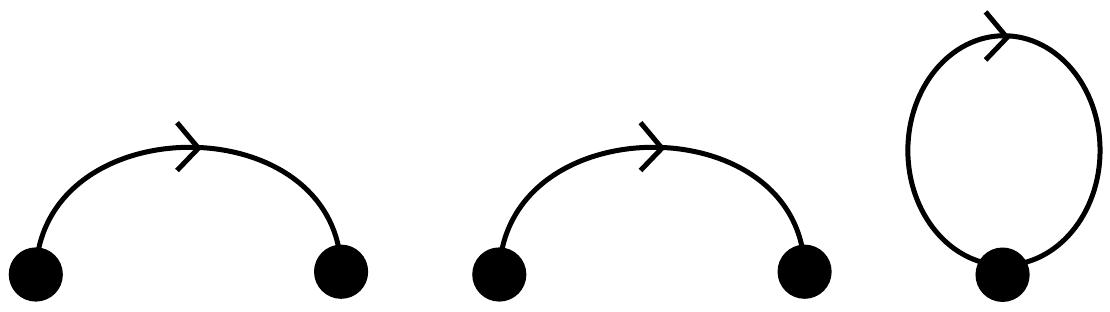}
	 	 \caption{$\sum_{i,j,k,l,m} M_{ij}M_{kl}M_{mm}$}
	 \label{fig:Graph7}
	\end{subfigure}%
	\vspace{5mm}
	\begin{subfigure}[b]{.35\textwidth}
	  	\centering
 		 \includegraphics[width=.8\linewidth]{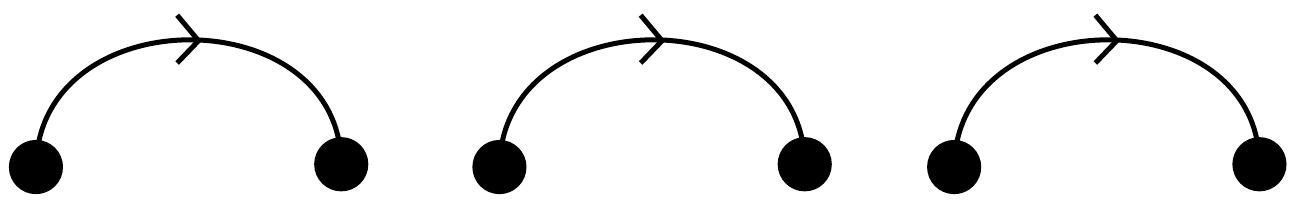}
	 	 \caption{$\sum_{i,j,k,l,m,n} M_{ij}M_{kl}M_{mn}$ }
	 \label{fig:Graph8}
	\end{subfigure}%
		\begin{subfigure}[b]{.35\textwidth}
	  	\centering
 		 \includegraphics[width=.81\linewidth]{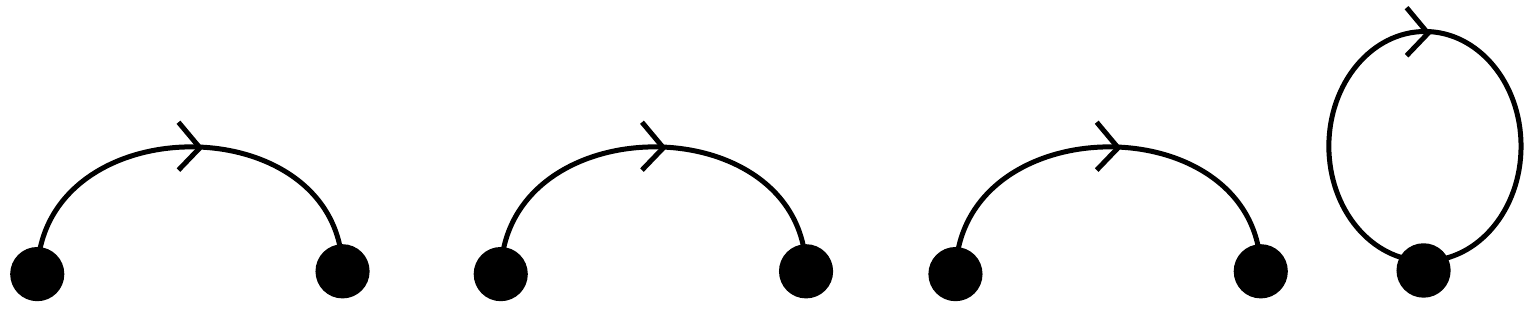}
	 	 \caption{$\sum_{i_{1}, \dots ,i_{7}}  M_{i_{1}i_{2}}M_{i_{3}i_{4}}M_{i_{5}i_{6}}M_{i_{7}i_{7}}$ }
	 \label{fig:Graph9}
	\end{subfigure}%
	\begin{subfigure}[b]{.35\textwidth}
	  	\centering
 		 \includegraphics[width=.8\linewidth]{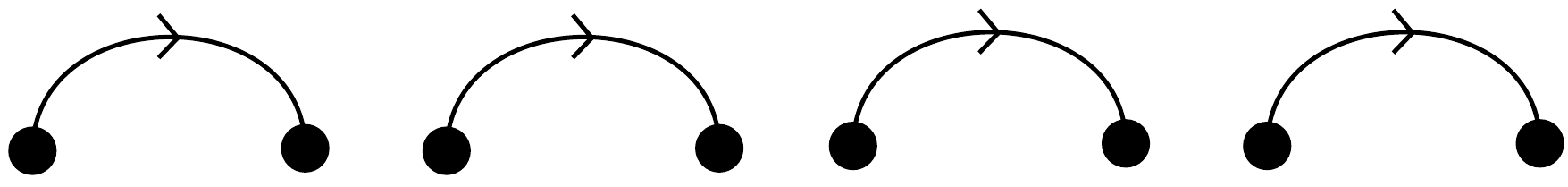}
	 	 \caption{$\sum_{i_{1}, \dots ,i_{8}} M_{i_{1}i_{2}}M_{i_{3}i_{4}}M_{i_{5}i_{6}}M_{i_{7}i_{8}}$ }
	 \label{fig:Graph10}
	\end{subfigure}%
	\vspace{3mm}
  \caption{The 10 higher order observable graph diagrams labelled with the associated sum.}
\end{figure}

\end{document}